\documentclass[sigconf,screen]{acmart}
\usepackage{siunitx}
\usepackage{algorithm}
\usepackage[noend]{algpseudocode}
\usepackage{booktabs}
\usepackage[textsize=small,bordercolor=white]{todonotes}

\usepackage{tabularx}

\newtheorem{guideline}{Guideline}

\AtBeginDocument{%
  \providecommand\BibTeX{{%
    \normalfont B\kern-0.5em{\scshape i\kern-0.25em b}\kern-0.8em\TeX}}}

\setcopyright{acmcopyright}
\copyrightyear{2018}
\acmYear{2018}
\acmDOI{XXXXXXX.XXXXXXX}

\acmConference[Conference acronym 'XX]{Make sure to enter the correct
  conference title from your rights confirmation emai}{June 03--05,
  2018}{Woodstock, NY}
\acmBooktitle{Woodstock '18: ACM Symposium on Neural Gaze Detection,
 June 03--05, 2018, Woodstock, NY} 
\acmPrice{15.00}
\acmISBN{978-1-4503-XXXX-X/18/06}

\begin{document}

\title{Where Do We Go From Here? Guidelines For Offline Recommender Evaluation}

\author{Tobias Schnabel}
\affiliation{%
  \institution{Microsoft Research}
  \city{Redmond}
  \country{USA}}
\email{toschnab@microsoft.com}

\begin{abstract}
  Various studies in recent years have pointed out large issues in the offline evaluation of recommender systems~\cite{cremonesi2021aimag,dacrema2019we}, making it difficult to assess whether true progress has been made. However, there has been little research into what set of practices should serve as a starting point during experimentation. In this paper, we examine four larger issues in recommender system research regarding uncertainty estimation, generalization, hyperparameter optimization and dataset pre-processing in more detail to arrive at a set of guidelines. We present a \textsc{TrainRec}, a lightweight and flexible toolkit for offline training and evaluation of recommender systems that implements these guidelines. Different from other frameworks, \textsc{TrainRec} is a toolkit that focuses on experimentation alone, offering flexible modules that can be can be used together or in isolation.
  
  Finally, we demonstrate \textsc{TrainRec}'s usefulness by evaluating a diverse set of twelve baselines across ten datasets. Our results show that (i) many results on smaller datasets are likely not statistically significant, (ii) there are at least three baselines that perform well on most datasets and should be considered in future experiments, and (iii) improved uncertainty quantification (via nested CV and statistical testing) rules out some reported differences between linear and neural methods. Given these results, we advocate that future research should standardize evaluation using our suggested guidelines.
\end{abstract}

\ccsdesc[500]{Information systems~Recommender systems}
\ccsdesc[300]{Information systems~Evaluation of retrieval results}
\ccsdesc[300]{Mathematics of computing~Resampling methods}

\keywords{evaluation, recommender systems, statistical testing, benchmark}

\maketitle

\section{Introduction}
Despite growing work that tests recommender systems online, offline evaluation is still by far the most popular evaluation paradigm used in recent research publications~\cite{sun2020we}.
What has been troubling is that an increasing amount of research has pointed out important issues with common protocols for offline evaluation of recommender systems~\cite{bellogin2011precision,ekstrand2011rethinking,konstan2013toward,dacrema2019we} even leading some researchers to publicly call it a community-wide crisis~\cite{cremonesi2021aimag}. The cumulative effect of these issues became widely visible when \citet{dacrema2019we} performed a series of reproducibility experiments showing that reported gains vanished in most cases when baselines were tuned properly. 
In a similar vein, \citet{rendle2019difficulty} showed that proper hyperparameter selection makes traditional matrix factorization-based approaches competitive to more recent methods. Overall, these discoveries mirror findings in information retrieval over a decade prior to that in 2009 when they compared all new published models of the last 15 years and found no overall improvements~\cite{armstrong2009improvements}. 

Besides a lack of properly tuned baselines, there are also doubts about the statistical reliability of results. A recent  survey~\cite{ihemelandu2021statistical} showed that 59\% of RecSys papers from the previous two years used no statistical testing at all, calling statistical inference the ``missing piece of RecSys experiment reliability discourse''. How to set up baselines and perform statistical analysis are just two areas of many where experimenters need to make design decisions. Another area pointed out by~\citet{sun2020we} is dataset pre-processing, e.g., how to deal with infrequent users or items.

With all of these design choices, there has been little research on how to guide them in a principled way. In this paper, we study four common issues that arise in almost all offline experimentation setups -- uncertainty estimation, generalization, hyperparameter optimization and dataset pre-processing. For each of these, we empirically examine different design choices to arrive at a set of guidelines. Even though the issues are very general, we study them in the specific and very common context of top-n recommendation from implicit feedback in this paper.

The contributions of this paper are three-fold. First, we empirically study four important issues for offline experimentation to come up with a set of informed guidelines. Second, we implement these in  a lightweight and flexible toolkit called \textsc{TrainRec} which is designed to accelerate and support all stages during offline experimentation. \textsc{TrainRec} provides a set of modules that can be used independently, allowing for maximum flexibility to make integration into existing code as painless as possible. 
Finally, we demonstrate \textsc{TrainRec}'s usefulness in a benchmark of a large and diverse set of baselines against a comprehensive set of commonly used recommender systems datasets. This, to the best of our knowledge, is the first consistent offline benchmark of its size comparing most recommender model classes. 
Our results show that many results on smaller datasets are most likely not statistically different, but there are three baselines (SLIM, MultiDAE, Ease) that perform well on most datasets, and should be considered for baselines in future experiments. Furthermore, \textsc{TrainRec}'s improved uncertainty estimation shows that some reported differences between linear and neural models are unlikely to exist.

Overall, we hope that \textsc{TrainRec} gives researchers and practitioners better tools to draw reliable conclusions when trying out new algorithms. We will make the complete code to reproduce all experiments public at \url{https://<OMITTED>}.


\section{Related Work}
This paper is a continuation of a longer, but recently accelerated trend in recommender systems research focused on research methodology and best practices. Around a decade ago, researchers started noticing that many recommender system papers had results that were difficult to compare due to widely differing implementations of baselines~\cite{said2014comparative}, diverging dataset processing choices~\cite{konstan2013toward}, and different evaluation protocols~\cite{ekstrand2011rethinking}. These findings motivated a push towards open-source recommender frameworks such as MyMediaLite~\cite{gantner2011mymedialite} or LensKit~\cite{ekstrand2011rethinking} which were among the first larger wave of  efforts. Nowadays, there are tens of such frameworks available, with RecBole~\cite{recbole} and Elliot~\cite{anelli2021elliot} being recent additions. However, having such a wide range of frameworks being available introduces new implementation differences on its own, and the bulk of replicability issues still remains unresolved. In this work, we argue that it is vital to report also how robust the outcomes of an experiment are. Without it, one cannot adequately replicate previous findings as they could have merely been to chance.

Lately, as a consequence of hugely varying evaluation protocols and baseline choices, a number of surveys have brought attention to reproducibility and replicability in the recommender community. For example, \citet{dacrema2019we} discovered that recently reported state-of-the-art results were often surpassed when the set of baselines was augmented with a wider set of adequate baselines. This has also been found to be the case in session-based recommendation~\cite{ludewig2019performance}. 
Since then, there has been more analysis of how consistent or diverging various elements of recommender system evaluation have been implemented in published work. Looking at the implementation of offline evaluation metrics, \citet{qualitymetrics} uncovered a number of troubling findings. First, there is considerable disagreement about how metrics should be implemented and Precision was the only metric that was consistently implemented. Second, only about half of all papers surveyed even specified how a certain metric was implemented. Another important factor during metric computation is what candidate set is being sampled at test time. Early work reported only smaller inconsistencies between sampling methods for precision~\cite{bellogin2011precision}, later work showed~\cite{krichene2020sampled,canamares2020target} that sampling leads to inconsistent orderings of methods and that adjustments are needed~\cite{li2020sampling}. Finally, a survey by \citet{ihemelandu2021statistical} showed that statistical testing is still missing in a lot of recently published work, making it hard to get insight into how much uncertainty was involved in the experiments. 

While there has been some research on what the possible impact of different design choices for the evaluation protocol is, most studies look only at individual factors such as sampling~\cite{bellogin2011precision,krichene2020sampled} or metric implementation~\cite{qualitymetrics}. Perhaps the most comprehensive study of design choices was carried out by~\citet{canamares2020offline} where they looked at how the ordering of two methods (UserKNN and Matrix Factorization) on two datasets changes with different settings. As expected, choices like data splitting and score aggregation can change outcomes. On the more surprising side is the finding that even less obvious choices, e.g., changing the gain factor used in computing nDCG, can flip outcomes. Because it is impossible to resolve all these issues in a single paper, we focus on the four groups of issues in this work that we believe are most important to robust experimentation.

\section{Issues in Offline Evaluation of Recommender Systems}
\label{sec:issues}
In this section, we study four larger issues regarding uncertainty estimation, generalization, hyperparameter optimization and dataset processing. While there exists a much wider range of open issues, e.g., biases in the data, we chose the set below because they represent fundamental and critical building blocks of most experimentation pipelines and are thus are most likely to provide practically relevant insights.

\subsection{Issue 1: Uncertainty Estimation}
Quantifying the uncertainty involved in an experiment is crucial for fair comparison and replicability. 
To illustrate this, assume we would like to compare two baselines, UserKNN and ItemKNN on the MovieLens 1M dataset~\cite{harper2015movielens} and decide which one performs better. We divide the dataset up into five equal-sized folds, and use the last four of them for training (and hyperparameter optimization) and first one for testing. The settings we use in this experiment were identical to those of Section~\ref{sec:benchmark}, but are not relevant for this exposition.

Running the two algorithms on this particular split which we call split A results in HitRate@50 scores of 0.428 for ItemKNN and 0.457 for UserKNN. Using the two-sided McNemar's test on the scores of each user, we find that UserKNN has a statistically significant higher score (p < $0.01$) and are inclined to conclude that UserKNN is the best-performing method. However, if we repeat this experiment on the other four possible splits (e.g., test on fold 2), and compare those scores (Table~\ref{tbl:sig}), we see that the overall variance between different dataset splits is considerable and it might be unwarranted to make such a claim. In fact, as the 95\% bootstrapped confidence intervals overlap ([0.432, 0.446] vs [0.441, 0.458]), we can now see that any superiority claims would be unwarranted.

\label{sec:uncertainty}
\begin{table}[!tb]
    \centering
    \caption{Even though HitRate@50 scores vary greatly for different dataset splits on ML1M, testing only on the first datasplit could find methods to be significantly different ($p < 0.01$). The NestedCV procedure of Algorithm~\ref{alg:crosscv} guards against this erroneous conclusion.}
    \label{tbl:sig}
    \begin{tabular}{rrrr}
    \hline
        \toprule
        ItemKNN & UserKNN & Dataset Split & significant?\\ 
        \midrule
        0.428 & 0.457 & A & \checkmark\\ 
        0.443 & 0.463 & B  & \text{\sffamily x}\\ 
        0.438 & 0.437 & C & \text{\sffamily x}\\ 
        0.452 & 0.446 & D &\text{\sffamily x}\\ 
        0.434 & 0.440 & E &\text{\sffamily x}\\ 
        \bottomrule
    \end{tabular}
\end{table}

\begin{table*}[!tb]
    \centering
    \caption{Collaborative filtering datasets are not static. In almost all of them, a recommender systems sees new users on 94\% of all days, clearly demonstrating the need for any practical recommender system to accommodate new users.}
    \label{tbl:newusers}
\begin{tabular}{l 
S[table-alignment=right, table-auto-round=true, table-format=3.1] 
S[table-alignment=right, table-format=1.3, table-auto-round=true] 
S[table-alignment=right, table-auto-round=true, table-format=5.1] 
S[table-alignment=right, table-format=1.3, table-auto-round=true] 
S[table-alignment=right, table-auto-round=true, table-format=2.1] 
S[table-alignment=right, table-format=1.3, table-auto-round=true]}\toprule
 & \multicolumn{2}{c}{new users} & \multicolumn{2}{c}{new ratings} & \multicolumn{2}{c}{new items}\\
\cmidrule(lr){2-3} \cmidrule(lr){4-5} \cmidrule(lr){6-7}
{dataset} & {new users / day} & {\% of days} & {new ratings / day} & {\% of days} & {new items / day} & {\% of days} \\ \midrule
        ml100k & 4.355140186915888 & 0.9345794392523364 & 463.0140186915888 & 0.9906542056074766 & 5.347619047619047 & 0.6666666666666666\\
        ml1m & 24.19918699186992 & 1.0 & 1157.9858323494686 & 1.0 & 6.751091703056769 & 0.834061135371179\\
        amazon-e & 735.4529360210341 & 0.9614373356704645 & 1369.5561787905347 & 0.9614373356704645 & 83.3763365468887 & 0.950920245398773\\
        ml20m & 19.79382929642445 & 0.9783737024221453 & 2855.267607662394 & 0.9867492438427193 & 3.8028289755679383 & 0.6865266466638091\\
        netflix & 218.5467461044913 & 0.9793767186067828 & 45213.97197106691 & 0.9832730560578662 & 8.363075462743236 & 0.9639297579496915\\
\bottomrule \end{tabular}
\end{table*}

We argue that the issue with the false discovery in the above experiment lies in the fact was done at the wrong level. Specifically, we only considered the uncertainty we have on the sample level, e.g., tested whether the mean scores of two  models were statistically different but failed to take into account other sources of randomness that affect results which can be grouped into four categories. First, there is randomness coming from the way the data is split, and different splits can lead to considerable variance~\cite{dacrema2019we}. Second, many recommender algorithms use numeric methods with random components (e.g., weight initialization) that lead to different outcomes, especially for highly non-convex deep learning architectures~\cite{sutskever2013importance}. Third, there is variance coming from hyperparameter selection~\cite{rendle2019difficulty}. Finally, as previously discussed and perhaps most used in statistical testing of recommender systems, is uncertainty at the sample level, e.g., the question of whether the population means between two labeled test sets differ~\cite{shani2011evaluating}. 

\begin{guideline}
\label{guide:uncertain}
When reporting results, quantify the uncertainty stemming from randomness in all components, e.g., via confidence intervals. In a typical setup, there is uncertainty associated with the (i) dataset splits, (ii) hyperparameter optimization, (iii) model training, and (iv) individual predictions.   
\end{guideline}

\subsection{Issue 2: Generalization}
Perhaps the largest practical gap lies in how recommender systems are expected to deal with previously unseen data points, for example new users, items or ratings during offline vs. online evaluation. The overwhelming majority of recommender systems are tested on the same users and items that they were also trained on, a setup also known as \emph{weak generalization}. This essentially amounts to a closed-world assumption which does not reflect the reality of recommender systems at all. To demonstrate this, we examined a variety of public recommender datasets with timestamped information for each rating. Table~\ref{tbl:newusers} shows the number of events per day and the percentage of days with at least one event. A user was deemed a new user on the day they rated their first item.
All the datasets we examined show a great amount of dynamic changes -- there are at least four new users each day and we see at least one new user on more than 94\% of the days. 
From this, we argue that any practical recommender system needs to go beyond weak generalization.

To align this terminology better with known concepts in machine learning, we use the terms \emph{transductive learning} instead of weak generalization and \emph{user-inductive learning} to describe the setting where test users are unknown (someimes also called \emph{strong generalization}). Transductive learning is defined assumes that all test instances are known beforehand~\cite{vapnik1999nature}, whereas inductive learning treats them as unknown. 
As a final goal, we would like to evaluate recommender systems in a fully inductive setting with both unknown users and unknown items at test time. However, this is a very challenging scenario and there are only a few existing baselines that would support it, so we propose the intermediate goal of evaluating new systems at least in a partially inductive scenario.

\begin{guideline}[Generalization]
\label{guide:gen}
Recommender systems should be evaluated in an inductive setting (e.g., using strong generalization) where they need to make predictions for unseen users (or items) at test time.
\end{guideline}

Looking at Table~\ref{tbl:newusers} again, we can see that in the datasets studied new users are seen much more often on average than new items. Because of this, we focus mainly on the user-inductive setting in this paper but note that changes to the item inventory might be more frequent in other domains (e.g., news, auctions). 

\subsubsection{User Profile Partitioning}
Related to the question of generalization is the question of how to split interactions from each user profile for training and evaluation. The most prevalent scheme according to a recent survey~\cite{sun2020we} is split-by-ratio where a certain percentage of a users' interactions go into training (e.g., 80\%) and the remainder goes into validation and test. 
\begin{figure}[tb]
    \centering
    \caption{The longer a user's profile, the more weight this user receives during evaluation under a split-by-ratio protocol.}
    \includegraphics[width=1.0\linewidth]{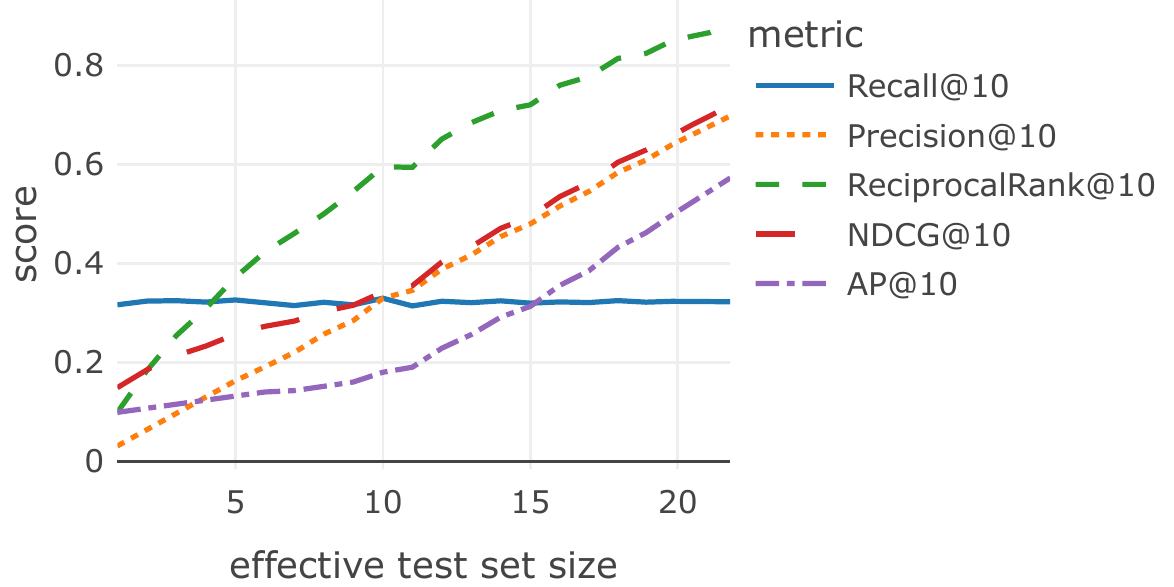}
    \label{fig:testsize}
\end{figure}

However, there are several issues with this scheme. A smaller one is that using relative proportions with a discrete number of ratings leads to further design choices -- i.e., how should one deal with users whose ratings cannot be split into the desired proportions?  
A far bigger issue of the split-by-ratio protocol is that it can confound results due to varying test set sizes. To illustrate this, we simulate a scenario where the recommender system produces recommendations of the same accuracy for each user and where we then gradually increase the size of each test set. More specifically, we draw a hidden set of positive items from a position-based relevance model that  sets an item at rank $r$ to relevant with probability $\propto -0.05\cdot e^{-r}$. We then randomly draw test items without replacement from this hidden set and plot how a set of common metrics change with the number of test items added. Note that we have to rely on synthetic data because in real-world datasets, we cannot ensure that each user receives recommendations of the same quality. As Figure~\ref{fig:testsize} shows, all metrics with the exception of Recall grow non-trivially as the test set size increases. For example, nDCG@10 grows about 4-fold when going from a test set size of one to 20. In other words, this means that a user with 20 items in their test set receives about 4 times the weight on average than a user with only one item. In essence, we argue that the split-by-ratio scheme is flawed it not gives more active users much higher weight and thus confounds testing. 
To circumvent the issues above, we recommend the following:
\begin{guideline}[User Profiles]
\label{guide:profile}
When partitioning user profiles, choose a hold-$n$ out protocol for validation and testing, i.e., hold $n$ items out each for validation and testing.
\end{guideline}
Of course, the larger $n$, the fewer user profiles have sufficient ratings to be split into multiple sets. Hence, unless there is a strong reason to do otherwise, we recommend to choose the most conservative setting of $n=1$, i.e., adopt a leave-out-one protocol. 


\subsection{Issue 3: Hyperparameter Optimization}
\begin{figure}[tb]
    \centering
    \caption{Bayesian Optimization finds the best hyperparameter settings the fastest. Random Search ends up finding similarly performing settings but shows less stable convergence.}
    \includegraphics[width=1.0\linewidth]{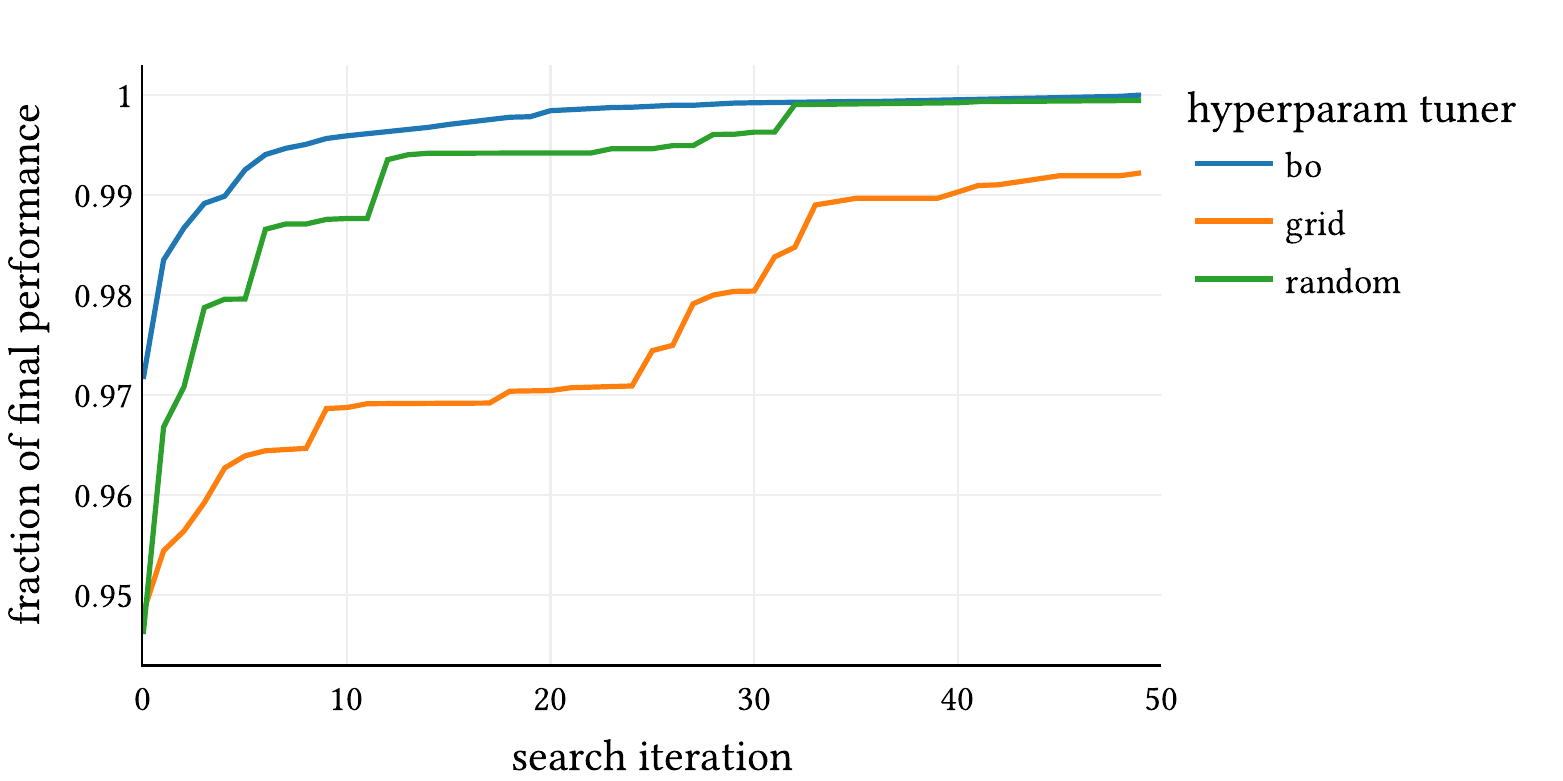}
    \label{fig:hyperparm_benchmark}
\end{figure}

The importance of proper hyperparameter tuning has been long known in the machine learning community, but only recently received more attention in the recommender systems community~\cite{dacrema2019we,rendle2019difficulty}. \citet{sun2020we} found that among their surveyed set of 85 recent recommender system papers, almost all of them use Grid Search even though better methods exist. To the best of our knowledge, there exists no empirical study of hyperparameter search algorithms in the recommender systems literature that could provide guidance on how to choose a set of reasonable defaults. 

To this end, we ran three different hyperparameter optimization algorithms with the datasets and baselines of Section~\ref{sec:benchmark} with a budget of 50 evaluations each. The first method was exhaustive search through all possible parameter combinations (Grid Search), where we discretized continuous parameters to make them fit the overall budget. Second was Random Search~\cite{bergstra2012random} which samples points at random from the overall search space. The third method was Bayesian Optimization which samples the next hyperparameter settings based on the uncertainty it has estimated in the search space.

Figure~\ref{fig:hyperparm_benchmark} shows the relative performance (best score so far over best score overall) of each hyperparameter tuning method, averaged across all datasets and baselines. Overall, we can see that Bayesian Optimization converges quickest. Grid search performs worst, both in terms of its final performance and in terms of its speed of convergence. Somewhat surprising is the competitive performance of Random Search relative to Bayesian Optimization. However, it shows less stable convergence and we recommend to only use it when parallel, independent hyperparameter optimization is needed.

\begin{guideline}[Hyperparameter Search Algorithms]
\label{guide:hyperopt}
Hyperparameter optimization should prefer Bayesian Optimization over Random Search and avoid Grid Search.
\end{guideline}
\begin{figure}[tbh]
    \centering
    \caption{Most methods are reaching 95\% of their overall performance increase after 40 search iterations with Bayesian Optimization.}
    \includegraphics[width=1.0\linewidth]{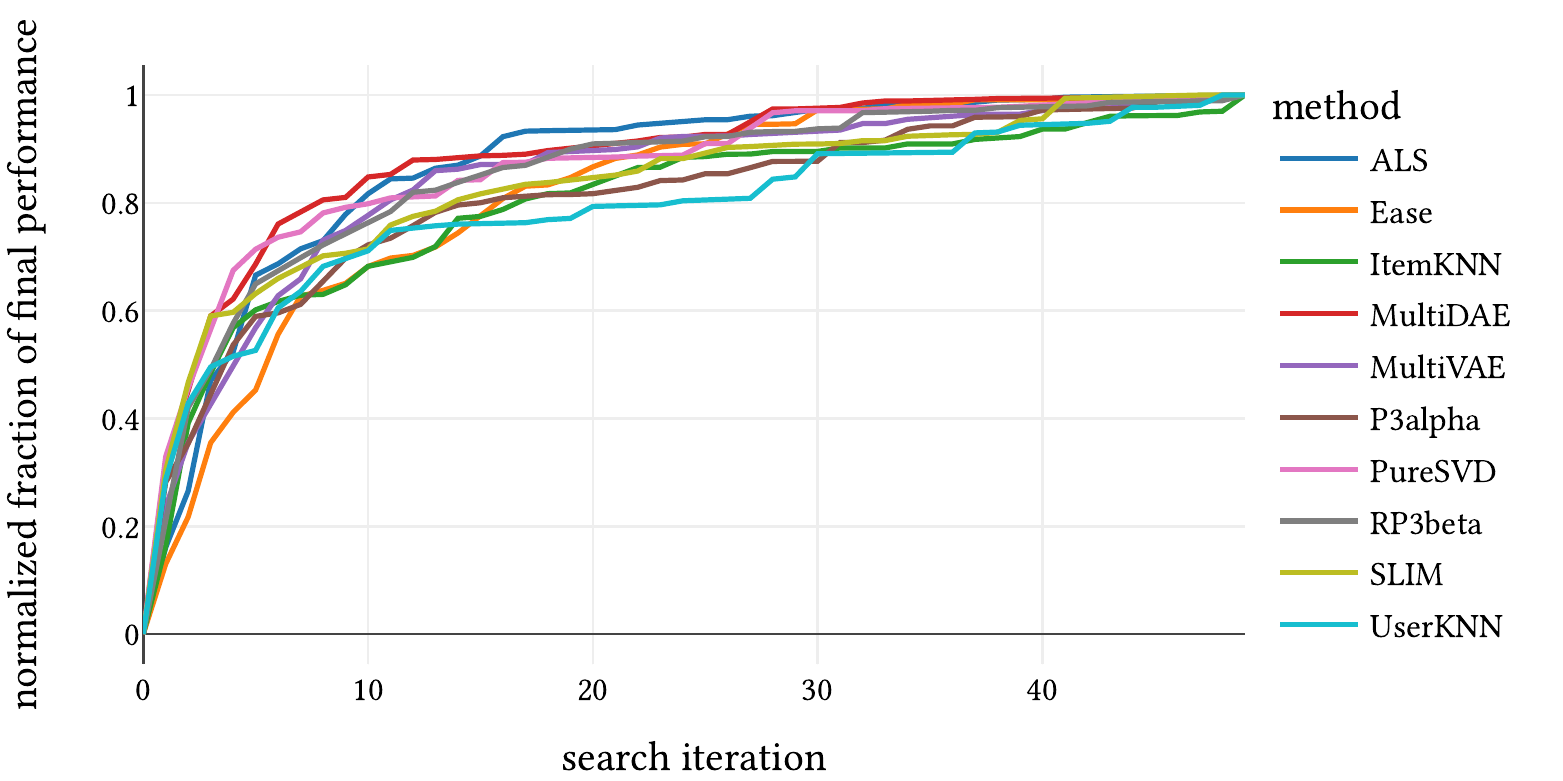}
    \label{fig:hyperparam_iters}
\end{figure}

To get a better picture of the number of iterations required for each baseline, Figure~\ref{fig:hyperparam_iters} shows the the normalized best scores found at each iteration. From it, we can see that most methods achieve 90-95\% of their final performance gain within the first 40 iterations under Bayesian optimization. 

\begin{guideline}[Hyperparameter Search Budget]
\label{guide:hpbudget}
Hyperparameter optimization with Bayesian Optimization should use a search budget of at least 40 iterations.
\end{guideline}

\subsection{Issue 4: Dataset Pre-Processing}
\begin{table}[tb]
    \centering
    \caption{When filtering out tail users and tail items, which filter is applied first matters as it changes the number of users and items (absolute differences shown).}
    \label{tbl:preprocess}
\begin{tabular}{l 
r 
r 
r}\toprule
{dataset} & {$\Delta$users} & {$\Delta$items} & {$\Delta$ratings} \\ \midrule
        bookx & 4873 & 3542 & 4873\\ 
        amazon-e & 22147 & 60335 & 22147\\ 
        yelp & 3525 & 33275 & 3525\\ 
\bottomrule \end{tabular}
\end{table}

We now turn to an issue regarding dataset handling that has received less attention in the literature. One of the most common pre-processing steps~\cite{sun2020we} is to filter out users or items that have an insufficient number of ratings. What has been less discussed is that when removing users (items) with less than $L$ interactions, the order in which these steps are carried out is critical as it can substantially change the data. To demonstrate this, we took three raw datasets, applied item and user filters with $L = 5$ in both orders, and recorded the absolute difference in counts in Table~\ref{tbl:preprocess}. We can see that the order of filtering can vastly impact the dataset characteristics -- for example, the amazon-e dataset can have about 22,000 fewer (or more) users depending on what filter was applied first. Note that interleaving those two steps via distributed processing could produce many more outcomes. 
To avoid this, we recommend the following:
\begin{guideline}[Filtering]
\label{guide:lcore}
When filtering out users or items with insufficient ratings, one should compute the $L$-core of the bipartite user-item graph where an edge indicates a positive rating. This corresponds to filtering repeatedly until all items and users meet the desired minimum.
\end{guideline}

The big advantage of this is that the $L$-core is unique~\cite{esfandiari2018parallel}, aiding reproducibility. Regarding the threshold $L$, a reasonable starting point is to choose it at least large enough that the training validation and test portions of each user contain at least rating each.

\section{TrainRec: A flexible library for offline experimentation}
\begin{figure}[tb]
    \centering
    \caption{TrainRec offers all the scaffolding needed to run reliable offline experiments while being highly flexible in how modules are picked.}
    \includegraphics[width=0.9\linewidth]{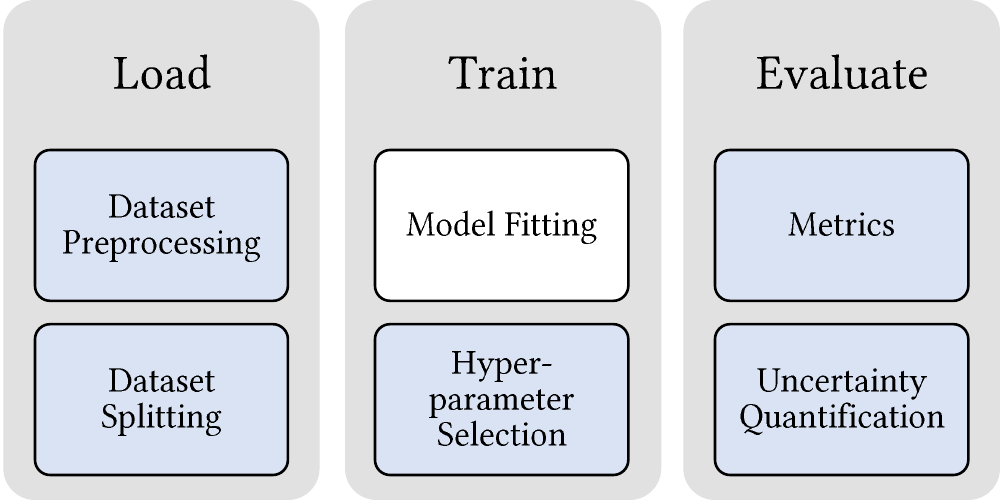}
    \label{fig:library}
\end{figure}

Based on our guidelines, we now introduce \textsc{TrainRec}, a highly flexible Python library for the offline training and evaluation of recommender systems. \textsc{TrainRec} provides an easy and highly flexible way for practitioners and researchers to implement the above mentioned guidelines in all stages during offline experimentation. Figure \ref{fig:library} shows an overview of the set of modules offered by \textsc{TrainRec} (light blue background). Only the model fitting itself needs to be provided; all other steps are implemented in \textsc{TrainRec}.

Different from previous frameworks, \textsc{TrainRec}'s purpose is not to provide a fixed experimentation workflow which often forces existing code to be re-written but to offer a set of loosely coupled modules that can be easily integrated into existing workflows. Modules can but do not have to be used together, allowing developers to opt in and out of functionality separately. Table~\ref{tbl:featurematrix} lists other common recommender frameworks which provide at least partial support for evaluation. Only \textsc{TrainRec} includes support for inductive generalization and uncertainty estimation.

\begin{table}[htp]
    \centering
    \caption{Only TrainRec provides support for important steps like inductive evaluation and uncertainty estimation.}
\begin{tabularx}{1.0\linewidth}{l
X
X
X
X
X}\toprule
{method} & {induct. eval} & de-\newline coupled & modern hyper-param search & uncer- tainty estim. & {metrics} \\ \midrule
        Cornac~\cite{cornac} &  &  &  &  & \checkmark\\ 
        Elliot~\cite{elliot} &  &  & \checkmark &  & \checkmark\\ 
        DaisyRec~\cite{sun2020we} &  &  & \checkmark &  & \checkmark\\ 
        RecBole~\cite{recbole} &  &  & \checkmark &  & \checkmark\\ 
        Recommenders~\cite{msftrecs} &  &  &  &  & \checkmark\\ 
        TrainRec & \checkmark & \checkmark & \checkmark & \checkmark & \checkmark\\
\bottomrule \end{tabularx}
    \label{tbl:featurematrix}
\end{table}

\subsection{Implementation of Guidelines}
We will now discuss how each of the guidelines discussed in Section~\ref{sec:issues} is implemented.

\subsubsection{Uncertainty Estimation}
To quantify uncertainty, \textsc{TrainRec} offers a procedure that combines nested cross-validation with bootstrapping. Algorithm~\ref{alg:crosscv} provides pseudo-code for the two steps. Intuitively, the \textsc{NestedCV} procedure~\cite{hastie2009elements}  corresponds to re-running the same experiment on $k$ different partitions, each time performing a full hyperparameter search over the new validation partition. Output is a list of $k$ scores that now adequately reflects the variability from hyperparameter selection, model fitting and dataset splitting. 

To make the resulting uncertainty estimates easily interpretable, we opted to compute confidence intervals via bootstrapping. Future work could examine the use of different tests, e.g., an Almost Stochastic Order test~\cite{dror2019deep}.
Another benefit of using this procedure is it does not rely on fixed data splits and thus the only parameter needed to replicate an experiment is the number of folds $k$.

\begin{algorithm}[t]
\caption{Bootstrapped Nested Cross-Validation}\label{alg:crosscv}
\begin{algorithmic}[1]
\Require $k$: number of outer cross-validation folds
\Procedure{BNCV}{data, k}
\State scores $\gets $ \Call{NestedCV}{data, k}
\State CI $\gets$ \Call{BootstrappedCI}{scores}
\State \textbf{return} \Call{Mean}{scores}, CI
\EndProcedure
\Procedure{NestedCV}{data, k}
\State scores $\gets []$ 
\For{$D_{\mathrm{train}}, D_{\mathrm{test}} \in$ \Call{CVFolds}{data, k}} \Comment{outer loop}
    \State best\_params $\gets $ \Call{OptimizeHyperparameters}{$D_{\mathrm{train}}$}
    \State chosen\_model $\gets$ \Call{FitModel}{$D_{\mathrm{train}}$, best\_params} 
    \State scores.append(\Call{Evaluate}{chosen\_model, $D_{\mathrm{test}}$})
\EndFor
\State \textbf{return} scores
\EndProcedure
\end{algorithmic}
\end{algorithm}

\subsubsection{Generalization}
\textsc{TrainRec} focuses on learning and evaluation in a partially inductive setting (e.g,. either generalizing with respect to new users or items). To do so, it offers data splitting routines that divide the set of users (or items) in non-overlapping partitions (Guideline~\ref{guide:gen}). A user (item) can then be further split via a hold-$n$ out protocol (Guideline~\ref{guide:lcore}).

\subsubsection{Hyperparameter Search}
\textsc{TrainRec} provides two competitive hyperparameter search algorithms. Following Guideline~\ref{guide:hyperopt}, both Bayesian Optimization and Random Search are provided through convenient high level interfaces and implemented via the scikit-optimize library~\cite{scikitopt}. We set the default number of search iterations to 50 to be conservative (Guideline~\ref{guide:hpbudget}).

\subsubsection{Dataset Pre-Processing}
\textsc{TrainRec} provides an efficient implementation for computing the $L$-core for user-item graphs (Guideline~\ref{guide:gen}). In addition, it contains routines to load common datasets into a sparse matrix format.


\section{Experiments and Results}
\label{sec:benchmark}
In this section, we demonstrate the usefulness of \textsc{TrainRec} by performing a comprehensive benchmark over ten datasets comparing twelve common baselines. We seek to answer the following two research questions:

\begin{description}
\item[RQ1] What does \textsc{TrainRec}'s uncertainty estimation reveal about the uncertainty of common methods on real-world data?
\item[RQ2] How do common baselines perform when evaluated fairly following Guidelines 1-6? 
\end{description}

\subsection{Experimental Setup}
\subsubsection{Baselines.} We considered the most commonly used baselines that support user-inductive generalization (Guideline~\ref{guide:gen}). There were roughly five groups of models:
\begin{enumerate}
    \item Unpersonalized: Random and Popularity make recommendations independent of a user in the dataset based on random scores or sorted by their item popularity.
    \item Similarity-Based: Includes UserKNN, ItemKNN, and newer variants such as P3Alpha and RP3beta~\cite{paudel2016updatable}. 
    \item Factorization-based: PureSVD applied a vanilla singular value decomposition to the user-item matrix; ALS implemented via weight matrix factorization and a fold-in step~\cite{hu2008collaborative}. Note that even though popular, BPR is not a user-inductive method.  
    \item Deep learning-based: MultiVAE and MultiDAE~\cite{liang2018variational} are two popular autoencoder-based architectures that can easily process new users.
    \item Linear: This includes the popular SLIM model~\cite{ning2011slim} as well as a newer simplified version called Ease~\cite{steck2019embarrassingly} which conveniently comes with a closed-form solution.
\end{enumerate}

\subsubsection{Metrics.}
Because of its simple and consistent definition, we considered HitRate@$k$ with with $k=50$ as our primary metric (other values of $k$ will be reported but not optimized over). HitRate@$k$ simply measures whether a held-out item occurred in the top $k$ or not and is equivalent to Recall@$k$ when the test set size is one:
\begin{equation}
\mathrm{HitRate}@k(y_1, \ldots, y_N) = \sum_{i=1}^{\min(k,N)} \mathrm{rel}(y_i),
\end{equation}
where $y_i$ is the result at rank $i$ and rel($y_i$) returns the relevance of that document.

\subsubsection{Uncertainty Estimation.} To quantify uncertainty, we used Algorithm~\ref{alg:crosscv} with $k=5$ folds. We optimized hyperparameters inside the loop using the users from the first three folds as the new training set and the fourth one as validation. Each user profile that was not part of training was split into a fold-in, validation ($n=1$) and test portion ($n=1$) following Guideline~\ref{guide:profile}.

\subsubsection{Hyperparameter Search.} 
Hyperparameter search was done using Bayesian Optimization (Guideline~\ref{guide:hyperopt} with a budget of 50 search iterations following Guideline~\ref{guide:hpbudget}. Hyperparameter ranges and names can be found in the supplementary material. 

\subsubsection{Datasets.} For this benchmark, we used a set of 10 publicly available and commonly used datasets, adding four more datasets to the six datasets used by the largest study we know of~\cite{sun2020we}. 

\subsubsection{Dataset Pre-Processing}
We used $5$-core filtering (Guideline~\ref{guide:lcore}) after binarizing all datasets with a threshold of $\lceil \frac{4}{5} n \rceil$ given a rating scale with $n$ levels except for the \verb+kuai+ dataset~\cite{gao2022kuairec} for which we used the threshold proposed by the authors. 

Table~\ref{tbl:datasets} lists the ten datasets with their basic statistics. Among the datasets considered, \verb+netflix+ and \verb+ml100k+ had the most and least number of ratings respectively. Orig. sparsity describes the original sparsity level in the data and was highest for \verb+kuai+ which had ratings for almost 100\% of all items and \verb+jester+ in which users rated about 50\% of the items. 

\begin{table}[tb]
\caption{Dataset statistics after binarizing and $5$-core filtering. $r_i$ and $r_u$ describe the ratings per item (user).}
\label{tbl:datasets}
\resizebox{\linewidth}{!}{
\begin{tabular}{l 
r 
r 
S[table-format=1.2e1, table-alignment=right, table-auto-round=true] 
S[table-format=1.2e1, table-alignment=right, table-auto-round=true] 
r
S[table-alignment=right, table-auto-round=true, table-format=5.1] 
S[table-alignment=right, table-auto-round=true, table-format=3.1]}\toprule
{dataset} & {users} & {items} & {orig. sparsity} & {5-core sp.} & {ratings} & {$r_i$} & {$r_u$} \\ \midrule
        ml100k & 938 & 1008 & 0.06304669364224531 & 0.05754920127254882 & {54.4\,K} & 53.98115079365079 & 58.00959488272921\\
        lastfm & 1859 & 2823 & 0.002782815119924182 & 0.013596719637756178 & {71.4\,K} & 25.276301806588734 & 38.383539537385694\\
        kuai & 1411 & 3065 & 0.9962024941648523 & 0.0501029547611808 & {216.7\,K} & 70.6952691680261 & 153.56555634301913\\
        bookx & 13854 & 34609 & 3.206771109452608e-05 & 0.0010868430683927898 & {521.1\,K} & 15.05712386951371 & 37.614551754006065\\
        ml1m & 6034 & 3125 & 0.044683625622312845 & 0.030460775604905534 & 574.4\,K & 183.80032 & 95.1899237653298\\
        jester & 50109 & 100 & 0.5633756009860938 & 0.20299746552515516 & 1.0\,M & 10172.0 & 20.299746552515515\\
        amazon-e & 124895 & 44843 & 3.912210290338533e-06 & 0.00019153791836615673 & 1.1\,M & 23.922128314341144 & 8.589134873293567\\
        yelp & 227847 & 90941 & 2.37365118038518e-05 & 0.00015343338412605032 & 3.2\,M & 34.95933627296819 & 13.953385385807142\\
        ml20m & 136674 & 13680 & 0.0053998478135544505 & 0.005336390486364146 & 10.0\,M & 729.3458333333333 & 73.00182185346152\\
        netflix & 463435 & 17721 & 0.01177557662406687 & 0.006925986106110141 & 56.9\,M & 3209.744371085153 & 122.73539978637781\\
\bottomrule \end{tabular}
}
\end{table}

\begin{table}[hp]
    \caption{HitRate@50 of 10 baselines across 12 datasets. }
    \vspace{-0.6em}
    \label{tab:results}
    \centering
    \setlength{\tabcolsep}{4pt}
    \footnotesize
    \begin{tabular}{l 
S[table-number-alignment=right, table-format = 1.3, round-mode=places, round-precision=3] 
 >{{[}}
        S[table-number-alignment=right, table-align-text-pre=false, table-format = -1.3, table-auto-round=true,table-space-text-pre={[}] @{,\,} 
        S[table-number-alignment=left, table-align-text-pre=false, table-format = -1.3, table-auto-round=true, table-space-text-post={]}] <{{]}} 
l 
S[table-number-alignment=right, table-format = 1.3, round-mode=places, round-precision=3] 
 >{{[}}
        S[table-number-alignment=right, table-align-text-pre=false, table-format = -1.3, table-auto-round=true,table-space-text-pre={[}] @{,\,} 
        S[table-number-alignment=left, table-align-text-pre=false, table-format = -1.3, table-auto-round=true, table-space-text-post={]}] <{{]}} 
} \multicolumn{4}{c}{ml100k} & \multicolumn{4}{c}{lastfm} \\ \cmidrule(r){0-3} \cmidrule(l){5-8}
{method} & {HR@50} & \multicolumn{2}{c}{95\% CI} & {method} & {HR@50} & \multicolumn{2}{c}{95\% CI}\\ 
 \cmidrule(r){0-3} \cmidrule(l){5-8}
        Random & 0.05013653430424392 & 0.039361702127659534 & 0.067379679144385 & Random & 0.01452337477900468 & 0.010752688172042961 & 0.01828681563921974\\
        Popularity & 0.3240869268403686 & 0.282586187279554 & 0.34928319490271925 & Popularity & 0.20601976639712488 & 0.18935309973045822 & 0.2247311827956989\\
        UserKNN & 0.5532711343725111 & 0.5213562407554898 & 0.5803390601888724 & PureSVD & 0.46424223980523427 & 0.4306451612903226 & 0.4784844796104686\\
        ItemKNN & 0.5692854704744568 & 0.556599158038457 & 0.5788883832062804 & UserKNN & 0.4674680462568472 & 0.4349462365591398 & 0.49031243659971596\\
        P3alpha & 0.5703208556149733 & 0.5335533052679485 & 0.590613266583229 & P3alpha & 0.49114424832623255 & 0.4559139784946237 & 0.5207141407993507\\
        RP3beta & 0.5714017521902377 & 0.5586187279553988 & 0.5931562179997725 & ALS & 0.49545401849114573 & 0.4489247311827957 & 0.5325507926846941\\
        PureSVD & 0.5767834793491865 & 0.5591534873136876 & 0.5918989646148594 & ItemKNN & 0.5045836593919368 & 0.4822580645161289 & 0.5244761325102164\\
        MultiVAE & 0.5778131755603596 & 0.5516497895096142 & 0.6012629423142565 & RP3beta & 0.5083499985508506 & 0.4639784946236559 & 0.5322913949511636\\
        MultiDAE & 0.5916657185117761 & 0.5773808169302537 & 0.6048867903060644 & Ease & 0.5217937570646031 & 0.485772251688259 & 0.5511143958496363\\
        ALS & 0.5970076231653203 & 0.5837581067243145 & 0.6070315166685629 & SLIM & 0.5303814161087441 & 0.4978494623655914 & 0.5594136741732603\\
        Ease & 0.6001706678803049 & 0.5811923995903971 & 0.6329787234042553 & MultiDAE & 0.5325536909833928 & 0.5 & 0.5519621482189955\\
        SLIM & 0.6055467061099101 & 0.5914893617021277 & 0.6183524860621231 & MultiVAE & 0.5379314842187636 & 0.517741935483871 & 0.5581210329536563\\
\end{tabular}
\begin{tabular}{l 
S[table-number-alignment=right, table-format = 1.3, round-mode=places, round-precision=3] 
 >{{[}}
        S[table-number-alignment=right, table-align-text-pre=false, table-format = -1.3, table-auto-round=true,table-space-text-pre={[}] @{,\,} 
        S[table-number-alignment=left, table-align-text-pre=false, table-format = -1.3, table-auto-round=true, table-space-text-post={]}] <{{]}} 
l 
S[table-number-alignment=right, table-format = 1.3, round-mode=places, round-precision=3] 
 >{{[}}
        S[table-number-alignment=right, table-align-text-pre=false, table-format = -1.3, table-auto-round=true,table-space-text-pre={[}] @{,\,} 
        S[table-number-alignment=left, table-align-text-pre=false, table-format = -1.3, table-auto-round=true, table-space-text-post={]}] <{{]}} 
}\multicolumn{4}{c}{kuai} & \multicolumn{4}{c}{bookx} \\ \cmidrule(r){0-3} \cmidrule(l){5-8}
{method} & {HR@50} & \multicolumn{2}{c}{95\% CI} & {method} & {HR@50} & \multicolumn{2}{c}{95\% CI}\\ 
 \cmidrule(r){0-3} \cmidrule(l){5-8}
        Random & 0.010630779640628479 & 0.00212014134275618 & 0.013467659073252581 & Random & 0.00158797863899824 & 0.0010827458710444197 & 0.0020933156323812404\\
        P3alpha & 0.3181966268200386 & 0.30141843971631205 & 0.3300102749166729 & Popularity & 0.05464143195317148 & 0.0513172140021652 & 0.056084954147325214\\
        MultiVAE & 0.31890584667819455 & 0.2950354609929078 & 0.33663634313209523 & PureSVD & 0.10942687739311352 & 0.10537712017322258 & 0.11411042944785274\\
        Popularity & 0.31890584667819455 & 0.29929078014184396 & 0.3326040648572789 & ALS & 0.11657254676139017 & 0.1106541057653599 & 0.122121977625406\\
        MultiDAE & 0.3203217803172694 & 0.29219858156028367 & 0.3387614966293261 & MultiVAE & 0.13043148024862972 & 0.12825694695055936 & 0.13201010465535906\\
        RP3beta & 0.32174022003358144 & 0.3035460992907801 & 0.33284214219482244 & MultiDAE & 0.13635023392094758 & 0.13123529281482915 & 0.141465175027066\\
        ItemKNN & 0.321745232188056 & 0.2985815602836879 & 0.33214294664561556 & UserKNN & 0.1370000273591751 & 0.13453626849512806 & 0.14023818116203532\\
        UserKNN & 0.3231611658271308 & 0.3078014184397163 & 0.33284715434929707 & ItemKNN & 0.1454454659984079 & 0.14133142774506977 & 0.14877617198238063\\
        PureSVD & 0.32740646066711776 & 0.31276595744680846 & 0.3453399493772397 & P3alpha & 0.1527356439242437 & 0.1517863587152652 & 0.15438469866474194\\
        Ease & 0.3295341202415858 & 0.30709219858156017 & 0.3448838433200511 & Ease & 0.15562292281976683 & 0.15122018012759794 & 0.15859337360777626\\
        SLIM & 0.3309475478034233 & 0.3156028368794326 & 0.3474525724882841 & RP3beta & 0.15692227518900626 & 0.1521586519482989 & 0.16023096355106453\\
        ALS & 0.3316567676615793 & 0.30921985815602837 & 0.348419918301882 & SLIM & 0.16089232601193118 & 0.15907614579574156 & 0.16234762567958236\\
\end{tabular}
\begin{tabular}{l 
S[table-number-alignment=right, table-format = 1.3, round-mode=places, round-precision=3] 
 >{{[}}
        S[table-number-alignment=right, table-align-text-pre=false, table-format = -1.3, table-auto-round=true,table-space-text-pre={[}] @{,\,} 
        S[table-number-alignment=left, table-align-text-pre=false, table-format = -1.3, table-auto-round=true, table-space-text-post={]}] <{{]}} 
l 
S[table-number-alignment=right, table-format = 1.3, round-mode=places, round-precision=3] 
 >{{[}}
        S[table-number-alignment=right, table-align-text-pre=false, table-format = -1.3, table-auto-round=true,table-space-text-pre={[}] @{,\,} 
        S[table-number-alignment=left, table-align-text-pre=false, table-format = -1.3, table-auto-round=true, table-space-text-post={]}] <{{]}} 
}\multicolumn{4}{c}{ml1m} & \multicolumn{4}{c}{jester} \\ \cmidrule(r){0-3} \cmidrule(l){5-8}
{method} & {HR@50} & \multicolumn{2}{c}{95\% CI} & {method} & {HR@50} & \multicolumn{2}{c}{95\% CI}\\ 
 \cmidrule(r){0-3} \cmidrule(l){5-8}
        Random & 0.01557731914852686 & 0.010610850744894661 & 0.0173985086992543 & Random & 0.6327205982583253 & 0.6298195641079496 & 0.636060666533626\\
        Popularity & 0.25024834402964463 & 0.23848363814729168 & 0.2586578293289147 & PureSVD & 0.8366560914555985 & 0.8335661544601877 & 0.8385352225104767\\
        ItemKNN & 0.44298639363250025 & 0.4302975594273867 & 0.45567522783761394 & Popularity & 0.8909177775165468 & 0.8857070377710698 & 0.8937138295749352\\
        P3alpha & 0.4547552214074614 & 0.44083641444805793 & 0.4647887323943662 & ItemKNN & 0.912191221424432 & 0.9034704072156912 & 0.9216703673034979\\
        UserKNN & 0.4584010354194231 & 0.4535792454463391 & 0.46313173156586573 & P3alpha & 0.9266598454958815 & 0.9235836075313484 & 0.9287766912791857\\
        RP3beta & 0.464864437821937 & 0.4564894390241556 & 0.4732394366197183 & ALS & 0.936558246640347 & 0.934341797611167 & 0.9387547395729395\\
        PureSVD & 0.4743098921300704 & 0.4545915822709155 & 0.4904722452361227 & MultiVAE & 0.9380949118804212 & 0.9367739799902542 & 0.939133905408102\\
        SLIM & 0.4961869745445652 & 0.4850041425020712 & 0.5050538525269264 & Ease & 0.9385539040933617 & 0.9362589529857985 & 0.9404709638794652\\
        MultiVAE & 0.502979716166475 & 0.4852498073015205 & 0.5186412593206297 & RP3beta & 0.9386137723831242 & 0.937174384401418 & 0.9403911394931151\\
        ALS & 0.5029815023199385 & 0.493618348467549 & 0.512178956089478 & SLIM & 0.938773457001522 & 0.9358810616643385 & 0.9404697252114602\\
        MultiDAE & 0.5096096430303605 & 0.4955232124382231 & 0.5260977630488815 & UserKNN & 0.9397912856360253 & 0.9382147540056123 & 0.9408090465619882\\
        Ease & 0.5165716570420474 & 0.5101093538108958 & 0.5221209610604806 & MultiDAE & 0.9469556139251853 & 0.946077545675335 & 0.9479944122929554\\
\end{tabular}
\begin{tabular}{l 
S[table-number-alignment=right, table-format = 1.3, round-mode=places, round-precision=3] 
 >{{[}}
        S[table-number-alignment=right, table-align-text-pre=false, table-format = -1.3, table-auto-round=true,table-space-text-pre={[}] @{,\,} 
        S[table-number-alignment=left, table-align-text-pre=false, table-format = -1.3, table-auto-round=true, table-space-text-post={]}] <{{]}} 
l 
S[table-number-alignment=right, table-format = 1.3, round-mode=places, round-precision=3] 
 >{{[}}
        S[table-number-alignment=right, table-align-text-pre=false, table-format = -1.3, table-auto-round=true,table-space-text-pre={[}] @{,\,} 
        S[table-number-alignment=left, table-align-text-pre=false, table-format = -1.3, table-auto-round=true, table-space-text-post={]}] <{{]}} 
}\multicolumn{4}{c}{amazon-e} & \multicolumn{4}{c}{yelp} \\ \cmidrule(r){0-3} \cmidrule(l){5-8}
{method} & {HR@50} & \multicolumn{2}{c}{95\% CI} & {method} & {HR@50} & \multicolumn{2}{c}{95\% CI}\\ 
 \cmidrule(r){0-3} \cmidrule(l){5-8}
        Random & 0.00119300212178224 & 0.00105688778573996 & 0.00128908282957678 & Random & 0.0006188368445331201 & 0.0005442273182984399 & 0.0007724526469210201\\
        Popularity & 0.0634773209495976 & 0.0623723928099603 & 0.06431002041715034 & Popularity & 0.0308277025949208 & 0.03037138841211518 & 0.031183071344546183\\
        PureSVD & 0.07018695704391684 & 0.06894591456823726 & 0.07164418111213414 & PureSVD & 0.11008702605163653 & 0.10907805395520334 & 0.1108063277312887\\
        ALS & 0.08901877577164814 & 0.08774570639337043 & 0.08970735417750904 & ALS & 0.1274714990060119 & 0.12689599438515065 & 0.12832621231956495\\
        MultiVAE & 0.11514472156611551 & 0.11279875095079861 & 0.11615356899795823 & UserKNN & 0.14192857028967917 & 0.14118808839342528 & 0.1423800053908848\\
        UserKNN & 0.11888386244445333 & 0.11692221466031465 & 0.1204211537691661 & RP3beta & 0.15132089473316673 & 0.13915801906782915 & 0.1554170598433145\\
        MultiDAE & 0.12636214420112896 & 0.125233195884543 & 0.12768325393330393 & ItemKNN & 0.15168951026783067 & 0.15119489126379768 & 0.15206390309201426\\
        ItemKNN & 0.12752311942031305 & 0.1253372833179871 & 0.12901237039112856 & P3alpha & 0.1535021308335954 & 0.15271279879300342 & 0.1541948959879129\\
        Ease & 0.12756315304856075 & 0.12529724968973935 & 0.12890027623203487 & SLIM & 0.15381373694016298 & 0.1530287960633798 & 0.15448894890040146\\
        RP3beta & 0.1287001080907962 & 0.12628207694463345 & 0.1305736818927899 & MultiVAE & 0.16754665231113514 & 0.1665013553804371 & 0.16904839989907824\\
        P3alpha & 0.12896433003723123 & 0.12665038632451256 & 0.1305416549901917 & MultiDAE & 0.1750428925823893 & 0.1743158726327108 & 0.17586646718668814\\
        SLIM & 0.1316065495015813 & 0.12887625605508618 & 0.1328155650746627 & Ease & \multicolumn{3}{c}{-- Out of memory --}\\
\end{tabular}
\begin{tabular}{l 
S[table-number-alignment=right, table-format = 1.3, round-mode=places, round-precision=3] 
 >{{[}}
        S[table-number-alignment=right, table-align-text-pre=false, table-format = -1.3, table-auto-round=true,table-space-text-pre={[}] @{,\,} 
        S[table-number-alignment=left, table-align-text-pre=false, table-format = -1.3, table-auto-round=true, table-space-text-post={]}] <{{]}} 
l 
S[table-number-alignment=right, table-format = 1.3, round-mode=places, round-precision=3] 
 >{{[}}
        S[table-number-alignment=right, table-align-text-pre=false, table-format = -1.3, table-auto-round=true,table-space-text-pre={[}] @{,\,} 
        S[table-number-alignment=left, table-align-text-pre=false, table-format = -1.3, table-auto-round=true, table-space-text-post={]}] <{{]}} 
}\multicolumn{4}{c}{ml20m} & \multicolumn{4}{c}{netflix} \\ \cmidrule(r){0-3} \cmidrule(l){5-8}
{method} & {HR@50} & \multicolumn{2}{c}{95\% CI} & {method} & {HR@50} & \multicolumn{2}{c}{95\% CI}\\ 
 \cmidrule(r){0-3} \cmidrule(l){5-8}
        Random & 0.00353395735769436 & 0.0031680995061276404 & 0.0037681159226322196 & Random & 0.00275766828142022 & 0.00262820028698734 & 0.0029022408752036\\
        Popularity & 0.24060904803693278 & 0.23885128955551488 & 0.24194067415729129 & Popularity & 0.17241900158598286 & 0.17125378963608698 & 0.17368131453170343\\
        P3alpha & 0.4317792759336438 & 0.42897384305835 & 0.433854033290653 & P3alpha & 0.36058778469472524 & 0.3597635051301692 & 0.36192562063719824\\
        ItemKNN & 0.461909306400808 & 0.4564870192572987 & 0.4668008048289738 & ItemKNN & 0.3736252117341159 & 0.3658679210676794 & 0.3791578106962142\\
        PureSVD & 0.46850168158086786 & 0.4654321175059831 & 0.47032010243277844 & UserKNN & 0.3865008037804654 & 0.38426100747677666 & 0.38871039088545317\\
        RP3beta & 0.47577440831824525 & 0.4687650796187757 & 0.48234863727821475 & RP3beta & 0.38889596167747364 & 0.3799842480606774 & 0.3969596599307346\\
        UserKNN & 0.4963927829533993 & 0.4877442899613502 & 0.5011670020120723 & PureSVD & 0.39844422626689824 & 0.39627132176033314 & 0.40061713077346345\\
        ALS & 0.5187599460147811 & 0.5162102157903085 & 0.5216316078287909 & MultiVAE & 0.41669058228230493 & 0.4147658247650695 & 0.41861533979954035\\
        MultiVAE & 0.5219792993846462 & 0.5202999817084324 & 0.523479207926808 & ALS & 0.4239580523698037 & 0.4222318124440321 & 0.42542751410661683\\
        SLIM & 0.5230255735708678 & 0.5213392393312362 & 0.5249387232485824 & SLIM & 0.4496035042670493 & 0.4477931101448963 & 0.45096076040868727\\
        Ease & 0.5483412722823167 & 0.5448924350228097 & 0.5499542710810317 & MultiDAE & 0.4606622287915242 & 0.4579606633076915 & 0.46253951471080085\\
        MultiDAE & 0.5639916667970467 & 0.5615259634862729 & 0.5662337662337663 & Ease & 0.4669004283232816 & 0.46507708740168524 & 0.46810879627132174\\ \midrule
\end{tabular}
\end{table}

\subsection{Results}
We provide the averaged results over five different test folds in Table~\ref{tab:results} along with their bootstrapped 95\% confidence intervals (CI)  for reference. Note that these results are for mainly reference and we will provide a summary graph below. Because Ease requires us to invert a matrix whose size is the number of items squared, we were unable to obtain results for the dataset with the largest set of items, \verb+yelp+, running out of memory. 

Two datasets that stand out from the rest are \verb+kuai+ and \verb+jester+ -- both were the results of attempts to create very dense or fully observed datasets~\cite{gao2022kuairec,goldberg2001eigentaste}. On both of these datasets, the Popularity baseline performs very competitively. For example, on \verb+kuai+, Popularity matches the performance of much more sophisticated MultiDAE method. On the \verb+jester+ dataset, Popularity even exceeds the performance of the typically robustly performing PureSVD baseline. Put another way, in these datasets most users like popular items (or vice versa). We will return to the discussion of the implications of item popularity in greater detail in the next section.

\subsubsection{RQ1: What does \textsc{TrainRec}'s uncertainty estimation reveal about real-world experiments?}
Starting with the smallest dataset we had in our benchmark, the \verb+ml100k+ dataset, we see that its results have wide confidence intervals. For example, RP3beta has confidence intervals for HitRate@50 overlap with those of seven other methods. The slightly larger datasets \verb+lastfm+ and \verb+kuai+ have similar levels of uncertainty associated, and even the \verb+ML1M+ dataset that is about six times as big as \verb+ml100k+ still has considerable uncertainty, e.g., the CI of the top five methods all overlap. It is also worth noting that even the largest datasets still have some uncertainty associated with them, albeit much smaller ranges. However, given that papers sometimes claim superiority based on a .005 difference, those differences are unlikely to hold up when an experiment is re-run with the same or slightly changed inputs. 

Note that the number of users is likely a big factor in the overall uncertainty because the individual estimates are averages over all test fold users. To see this, compare the confidence intervals of \verb+ml100k+ with those of \verb+jester+ -- even though \verb+jester+ has a small number of items it has sufficient users to average over.

Overall, these findings imply that caution needs to be exercised -- especially with small, but also with larger datasets due to the amounts of uncertainty involved. We will discuss further implications of this in the next section.

\subsubsection{RQ2: How do common baselines perform when following all guidelines?}
\begin{figure}
    \centering
    \caption{Among the baselines, SLIM, Ease and MultaiDAE belong to the top performing methods across datasets for our primary metric of HitRate@50, reaching on average 97\% of the performance that the best method per dataset yielded. Values for other cutoffs are reported for reference.}
    \includegraphics[width=0.9\linewidth]{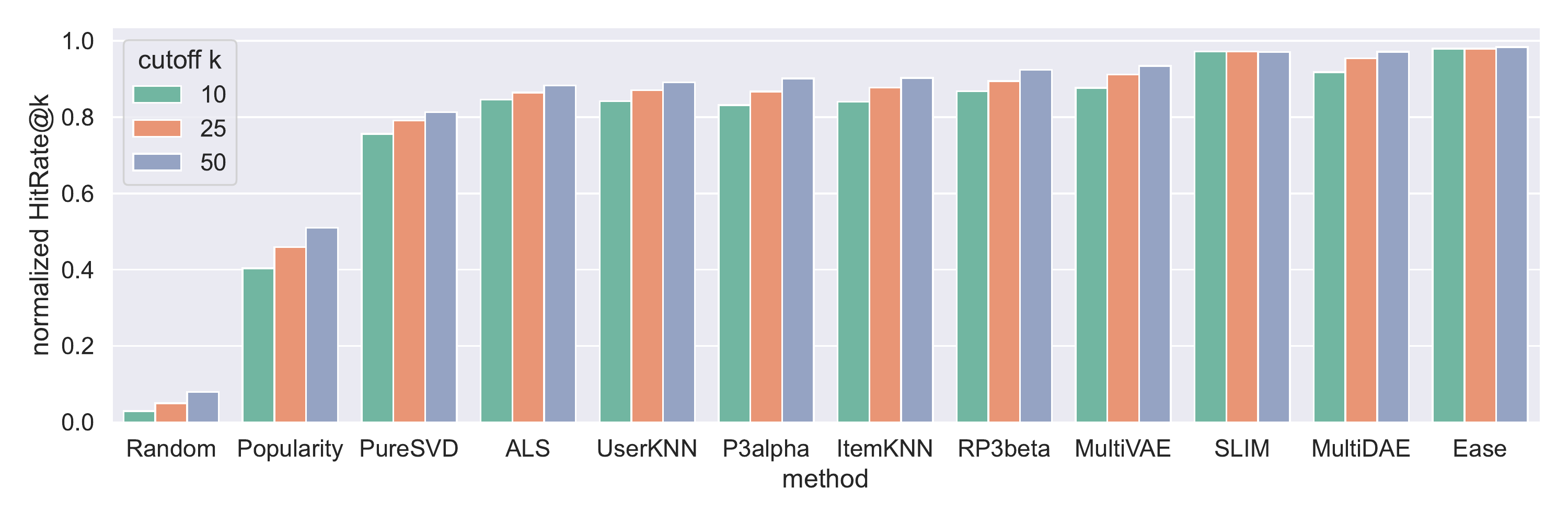}
    \label{fig:normalized_hitrate}
\end{figure}

Since metrics have different ranges on different datasets, we computed the \emph{normalized} scores for each dataset by expressing its performance relative to the \emph{best} performing method on that dataset. For example, on \verb+netflix+, ALS obtained a HitRate@50 of 0.424 and the best method was Ease with 0.467, resulting in a normalized HitRate@50 of 0.242/0.467 $\approx $ 0.906.

Figure~\ref{fig:normalized_hitrate} shows the normalized HitRate@50 of each method averaged over all ten datasets. Clearly, random is performing worst. Next comes Popularity with a normalized HitRate@50 of 0.51 -- meaning that it roughly reaches 50\% of the total performance the best methods would have gotten. The group of middle performers are ALS, ItemKNN, UserKNN, RP3beta and MultiVAE with scores ranging from 0.89-0.93. Finally, top performing with a average normalized score of 0.97 or higher were SLIM, MultiDAE and Ease. 

We also report normalized Recall@$k$ for smaller cutoff values. Even though one is tempted to compare performance within method for each $k$ this is not a valid comparison because the normalization differs between cutoffs. However, we can still compare methods using these secondary metrics (which were not optimized over) under a fixed cutoff $k$. When comparing the top three methods, MultiDAE appears to have lower relative recall performance at lower cutoff values ($k=10, 25$). This might be due to the smoothing properties of autoencoders, but further experiments are needed to verify this hypothesis.  

An interesting observation is that set of top performers consists of only one neural method and two linear models. The finding that linear models are highly competitive has also been noticed by others. For example, \citet{jin2021towards} also found only minute differences in performance among the SLIM, MultiDAE and Ease. They also tested variants of Ease but those failed to consistently improve over the SLIM, MultiDAE or Ease in its original formulation, again underlining how competitive they are. One difference though is that they found MultiVAE to perform slightly better than MultiDAE -- but they only reported results for one run, making it hard to generalize their conclusion. The similar performance of SLIM and Ease is very likely due to their close connection -- Ease essentially is the same as SLIM without an L1 sparsity constraint~\cite{steck2020admm}.

To answer our second research question, our large benchmark using 10 datasets and 12 baselines shows that among all baselines, SLIM, MultiDAE and Ease behaved most competitively. We recommend that those should be included in future experiments whenever possible. 

\section{Discussion}
We now discuss some of our findings in greater detail. When looking at the full set of results, we pointed out two potentially unwanted behaviors. First, smaller datasets had large confidence intervals associated with them. This is problematic when wanting to make superiority claims on such datasets, especially when only a single split is being used. A straightforward solution to this problem would be to drop such small datasets and only focus on larger ones. However, this may get rid of interesting dense datasets such as \verb+kuai+ which are mostly small in size (due to the high cost of labeling). Another potential approach could be to increase the number of outer folds -- in the extreme, leave-one-user-out --  so that the confidence interval will be computed over more samples. The second potential issue we uncovered was that on the two dense datasets, the Popularity baseline performed quite competitively. This finding points to a more fundamental question of what portion of the observed popularity of items is due to intrinsic factors, such as item qualities, and which portion is due to confounding factors such as perceived popularity from others~\cite{canamares2018should}. 

In our large-scale benchmark, we found that Ease, SLIM and MultiDAE were among the best performing baselines. This finding is somewhat surprising -- Ease is a simple linear model with a closed form solution whereas MultiDAE as a neural model needs to be trained via costly gradient descent. It also raises the question of what neural models are able to add beyond matching performance. We plan to study this question in more detail in future work.

To end our discussion, we would like to emphasize the list of issues we looked at is not exhaustive. In particular, offline evaluation has limitations on its own -- e.g., selection biases due to interfaces or user trust~\cite{marlin2003modeling}. However, our focus here was to provide guidance on what we believe are the most important and practical steps towards reliable offline experimentation. We would also like to emphasize that the guidelines we provide in this paper are starting points and acknowledge that there are circumstances in which there are good reasons to decide differently.

\section{Conclusions \& Future Work}
In this paper, we examine four larger issues that are part of most offline evaluation protocols for recommendation and derive a set six guidelines to serve as a starting point for practitioners and researchers. 
To implement these, we introduce \textsc{TrainRec}, a Python toolkit which supports all common stages during offline experimentation. Different from other frameworks, \textsc{TrainRec} provides uncertainty estimation and supports evaluation in partially inductive settings.

We show how \textsc{TrainRec} can reveal importance insights about the robustness of offline results. In a large-scale benchmark, we compare twelve common baselines on ten datasets and find considerable uncertainty about how well a single method does on a particular dataset. While the uncertainty is particularly large for smaller datasets, there is still enough residual uncertainty in larger datasets to call into question methods that claim superiority based on smaller differences.  Our benchmark also revealed that SLIM, Ease and MultiDAE were the baselines that overall perform the strongest and we recommend that those are included in future experiments. Interestingly, MultiDAE (a neural method) and Ease (a linear model) performed statistically indistinguishable on most datasets.  

We hope that the guidelines in this paper along with \textsc{TrainRec} will lead to improved overall replicability which in turn, can accelerate research in the longer term. There are many avenues to improve and extend our guidelines and \textsc{TrainRec} in the future. For example, one could add other algorithms to quantify uncertainty (e.g., through parametric bootstrapping) or add support for fully inductive evaluation.
Finally, we hope that this paper will enable practitioners to make more informed decisions about their evaluation setup and will spark more discourse around experimentation and evaluation of recommender systems in the future.

\bibliographystyle{ACM-Reference-Format}
\bibliography{references}


\begin{thebibliography}{42}


\ifx \showCODEN    \undefined \def \showCODEN     #1{\unskip}     \fi
\ifx \showDOI      \undefined \def \showDOI       #1{#1}\fi
\ifx \showISBNx    \undefined \def \showISBNx     #1{\unskip}     \fi
\ifx \showISBNxiii \undefined \def \showISBNxiii  #1{\unskip}     \fi
\ifx \showISSN     \undefined \def \showISSN      #1{\unskip}     \fi
\ifx \showLCCN     \undefined \def \showLCCN      #1{\unskip}     \fi
\ifx \shownote     \undefined \def \shownote      #1{#1}          \fi
\ifx \showarticletitle \undefined \def \showarticletitle #1{#1}   \fi
\ifx \showURL      \undefined \def \showURL       {\relax}        \fi
\providecommand\bibfield[2]{#2}
\providecommand\bibinfo[2]{#2}
\providecommand\natexlab[1]{#1}
\providecommand\showeprint[2][]{arXiv:#2}

\bibitem[Anelli et~al\mbox{.}(2021a)]%
        {anelli2021elliot}
\bibfield{author}{\bibinfo{person}{Vito~Walter Anelli},
  \bibinfo{person}{Alejandro Bellog{\'\i}n}, \bibinfo{person}{Antonio Ferrara},
  \bibinfo{person}{Daniele Malitesta}, \bibinfo{person}{Felice~Antonio Merra},
  \bibinfo{person}{Claudio Pomo}, \bibinfo{person}{Francesco~Maria Donini},
  {and} \bibinfo{person}{Tommaso Di~Noia}.} \bibinfo{year}{2021}\natexlab{a}.
\newblock \showarticletitle{Elliot: a comprehensive and rigorous framework for
  reproducible recommender systems evaluation}. In
  \bibinfo{booktitle}{\emph{Proceedings of the 44th International ACM SIGIR
  Conference on Research and Development in Information Retrieval}}.
  \bibinfo{pages}{2405--2414}.
\newblock


\bibitem[Anelli et~al\mbox{.}(2021b)]%
        {elliot}
\bibfield{author}{\bibinfo{person}{Vito~Walter Anelli},
  \bibinfo{person}{Alejandro Bellog{\'{\i}}n}, \bibinfo{person}{Antonio
  Ferrara}, \bibinfo{person}{Daniele Malitesta},
  \bibinfo{person}{Felice~Antonio Merra}, \bibinfo{person}{Claudio Pomo},
  \bibinfo{person}{Francesco~Maria Donini}, {and} \bibinfo{person}{Tommaso~Di
  Noia}.} \bibinfo{year}{2021}\natexlab{b}.
\newblock \showarticletitle{Elliot: {A} Comprehensive and Rigorous Framework
  for Reproducible Recommender Systems Evaluation}. In
  \bibinfo{booktitle}{\emph{{SIGIR} '21: The 44th International {ACM} {SIGIR}
  Conference on Research and Development in Information Retrieval, Virtual
  Event, Canada, July 11-15, 2021}}. \bibinfo{publisher}{{ACM}},
  \bibinfo{pages}{2405--2414}.
\newblock


\bibitem[Argyriou et~al\mbox{.}(2020)]%
        {msftrecs}
\bibfield{author}{\bibinfo{person}{Andreas Argyriou}, \bibinfo{person}{Miguel
  Gonz\'{a}lez-Fierro}, {and} \bibinfo{person}{Le Zhang}.}
  \bibinfo{year}{2020}\natexlab{}.
\newblock \showarticletitle{Microsoft Recommenders: Best Practices for
  Production-Ready Recommendation Systems}. In
  \bibinfo{booktitle}{\emph{Companion Proceedings of the Web Conference 2020}}
  \emph{(\bibinfo{series}{WWW '20})}. \bibinfo{pages}{50–51}.
\newblock


\bibitem[Armstrong et~al\mbox{.}(2009)]%
        {armstrong2009improvements}
\bibfield{author}{\bibinfo{person}{Timothy~G Armstrong},
  \bibinfo{person}{Alistair Moffat}, \bibinfo{person}{William Webber}, {and}
  \bibinfo{person}{Justin Zobel}.} \bibinfo{year}{2009}\natexlab{}.
\newblock \showarticletitle{Improvements that don't add up: ad-hoc retrieval
  results since 1998}. In \bibinfo{booktitle}{\emph{Proceedings of the 18th ACM
  conference on Information and knowledge management}}.
  \bibinfo{pages}{601--610}.
\newblock


\bibitem[Bellogin et~al\mbox{.}(2011)]%
        {bellogin2011precision}
\bibfield{author}{\bibinfo{person}{Alejandro Bellogin}, \bibinfo{person}{Pablo
  Castells}, {and} \bibinfo{person}{Ivan Cantador}.}
  \bibinfo{year}{2011}\natexlab{}.
\newblock \showarticletitle{Precision-oriented evaluation of recommender
  systems: an algorithmic comparison}. In \bibinfo{booktitle}{\emph{Proceedings
  of the fifth ACM conference on Recommender systems}}.
  \bibinfo{pages}{333--336}.
\newblock


\bibitem[Bergstra and Bengio(2012)]%
        {bergstra2012random}
\bibfield{author}{\bibinfo{person}{James Bergstra} {and}
  \bibinfo{person}{Yoshua Bengio}.} \bibinfo{year}{2012}\natexlab{}.
\newblock \showarticletitle{Random search for hyper-parameter optimization.}
\newblock \bibinfo{journal}{\emph{Journal of machine learning research}}
  \bibinfo{volume}{13}, \bibinfo{number}{2} (\bibinfo{year}{2012}).
\newblock


\bibitem[Ca{\~n}amares and Castells(2018)]%
        {canamares2018should}
\bibfield{author}{\bibinfo{person}{Roc{\'\i}o Ca{\~n}amares} {and}
  \bibinfo{person}{Pablo Castells}.} \bibinfo{year}{2018}\natexlab{}.
\newblock \showarticletitle{Should I follow the crowd? A probabilistic analysis
  of the effectiveness of popularity in recommender systems}. In
  \bibinfo{booktitle}{\emph{The 41st International ACM SIGIR Conference on
  Research \& Development in Information Retrieval}}.
  \bibinfo{pages}{415--424}.
\newblock


\bibitem[Ca{\~n}amares and Castells(2020)]%
        {canamares2020target}
\bibfield{author}{\bibinfo{person}{Roc{\'\i}o Ca{\~n}amares} {and}
  \bibinfo{person}{Pablo Castells}.} \bibinfo{year}{2020}\natexlab{}.
\newblock \showarticletitle{On target item sampling in offline recommender
  system evaluation}. In \bibinfo{booktitle}{\emph{Fourteenth ACM Conference on
  Recommender Systems}}. \bibinfo{pages}{259--268}.
\newblock


\bibitem[Ca{\~n}amares et~al\mbox{.}(2020)]%
        {canamares2020offline}
\bibfield{author}{\bibinfo{person}{Roc{\'\i}o Ca{\~n}amares},
  \bibinfo{person}{Pablo Castells}, {and} \bibinfo{person}{Alistair Moffat}.}
  \bibinfo{year}{2020}\natexlab{}.
\newblock \showarticletitle{Offline evaluation options for recommender
  systems}.
\newblock \bibinfo{journal}{\emph{Information Retrieval Journal}}
  \bibinfo{volume}{23}, \bibinfo{number}{4} (\bibinfo{year}{2020}),
  \bibinfo{pages}{387--410}.
\newblock


\bibitem[contributors(2022)]%
        {scikitopt}
\bibfield{author}{\bibinfo{person}{Various contributors}.}
  \bibinfo{year}{2022}\natexlab{}.
\newblock \bibinfo{title}{Scikit-Optimize}.
\newblock
\newblock
\urldef\tempurl%
\url{https://scikit-optimize.github.io/stable/index.html}
\showURL{%
\tempurl}


\bibitem[Cremonesi and Jannach(2021)]%
        {cremonesi2021aimag}
\bibfield{author}{\bibinfo{person}{Paolo Cremonesi} {and}
  \bibinfo{person}{Dietmar Jannach}.} \bibinfo{year}{2021}\natexlab{}.
\newblock \showarticletitle{Progress in recommender systems research: Crisis?
  What crisis?}
\newblock \bibinfo{journal}{\emph{AI Magazine}} \bibinfo{volume}{42},
  \bibinfo{number}{3} (\bibinfo{year}{2021}), \bibinfo{pages}{43--54}.
\newblock


\bibitem[Dacrema et~al\mbox{.}(2019)]%
        {dacrema2019we}
\bibfield{author}{\bibinfo{person}{Maurizio~Ferrari Dacrema},
  \bibinfo{person}{Paolo Cremonesi}, {and} \bibinfo{person}{Dietmar Jannach}.}
  \bibinfo{year}{2019}\natexlab{}.
\newblock \showarticletitle{Are we really making much progress? A worrying
  analysis of recent neural recommendation approaches}. In
  \bibinfo{booktitle}{\emph{Proceedings of the 13th ACM conference on
  recommender systems}}. \bibinfo{pages}{101--109}.
\newblock


\bibitem[Dror et~al\mbox{.}(2019)]%
        {dror2019deep}
\bibfield{author}{\bibinfo{person}{Rotem Dror}, \bibinfo{person}{Segev
  Shlomov}, {and} \bibinfo{person}{Roi Reichart}.}
  \bibinfo{year}{2019}\natexlab{}.
\newblock \showarticletitle{Deep Dominance - How to Properly Compare Deep
  Neural Models}. In \bibinfo{booktitle}{\emph{Proceedings of the 57th
  Conference of the Association for Computational Linguistics, {ACL} 2019,
  Florence, Italy, July 28- August 2, 2019, Volume 1: Long Papers}},
  \bibfield{editor}{\bibinfo{person}{Anna Korhonen}, \bibinfo{person}{David~R.
  Traum}, {and} \bibinfo{person}{Llu{\'{\i}}s M{\`{a}}rquez}} (Eds.).
  \bibinfo{publisher}{Association for Computational Linguistics},
  \bibinfo{pages}{2773--2785}.
\newblock
\urldef\tempurl%
\url{https://doi.org/10.18653/v1/p19-1266}
\showDOI{\tempurl}


\bibitem[Ekstrand et~al\mbox{.}(2011)]%
        {ekstrand2011rethinking}
\bibfield{author}{\bibinfo{person}{Michael~D Ekstrand},
  \bibinfo{person}{Michael Ludwig}, \bibinfo{person}{Joseph~A Konstan}, {and}
  \bibinfo{person}{John~T Riedl}.} \bibinfo{year}{2011}\natexlab{}.
\newblock \showarticletitle{Rethinking the recommender research ecosystem:
  reproducibility, openness, and lenskit}. In
  \bibinfo{booktitle}{\emph{Proceedings of the fifth ACM conference on
  Recommender systems}}. \bibinfo{pages}{133--140}.
\newblock


\bibitem[Esfandiari et~al\mbox{.}(2018)]%
        {esfandiari2018parallel}
\bibfield{author}{\bibinfo{person}{Hossein Esfandiari}, \bibinfo{person}{Silvio
  Lattanzi}, {and} \bibinfo{person}{Vahab Mirrokni}.}
  \bibinfo{year}{2018}\natexlab{}.
\newblock \showarticletitle{Parallel and streaming algorithms for k-core
  decomposition}. In \bibinfo{booktitle}{\emph{International Conference on
  Machine Learning}}. PMLR, \bibinfo{pages}{1397--1406}.
\newblock


\bibitem[Gantner et~al\mbox{.}(2011)]%
        {gantner2011mymedialite}
\bibfield{author}{\bibinfo{person}{Zeno Gantner}, \bibinfo{person}{Steffen
  Rendle}, \bibinfo{person}{Christoph Freudenthaler}, {and}
  \bibinfo{person}{Lars Schmidt-Thieme}.} \bibinfo{year}{2011}\natexlab{}.
\newblock \showarticletitle{MyMediaLite: A free recommender system library}. In
  \bibinfo{booktitle}{\emph{Proceedings of the fifth ACM conference on
  Recommender systems}}. \bibinfo{pages}{305--308}.
\newblock


\bibitem[Gao et~al\mbox{.}(2022)]%
        {gao2022kuairec}
\bibfield{author}{\bibinfo{person}{Chongming Gao}, \bibinfo{person}{Shijun Li},
  \bibinfo{person}{Wenqiang Lei}, \bibinfo{person}{Biao Li},
  \bibinfo{person}{Peng Jiang}, \bibinfo{person}{Jiawei Chen},
  \bibinfo{person}{Xiangnan He}, \bibinfo{person}{Jiaxin Mao}, {and}
  \bibinfo{person}{Tat-Seng Chua}.} \bibinfo{year}{2022}\natexlab{}.
\newblock \showarticletitle{KuaiRec: A Fully-observed Dataset for Recommender
  Systems}.
\newblock \bibinfo{journal}{\emph{arXiv preprint arXiv:2202.10842}}
  (\bibinfo{year}{2022}).
\newblock


\bibitem[Goldberg et~al\mbox{.}(2001)]%
        {goldberg2001eigentaste}
\bibfield{author}{\bibinfo{person}{Ken Goldberg}, \bibinfo{person}{Theresa
  Roeder}, \bibinfo{person}{Dhruv Gupta}, {and} \bibinfo{person}{Chris
  Perkins}.} \bibinfo{year}{2001}\natexlab{}.
\newblock \showarticletitle{Eigentaste: A constant time collaborative filtering
  algorithm}.
\newblock \bibinfo{journal}{\emph{information retrieval}} \bibinfo{volume}{4},
  \bibinfo{number}{2} (\bibinfo{year}{2001}), \bibinfo{pages}{133--151}.
\newblock


\bibitem[Harper and Konstan(2015)]%
        {harper2015movielens}
\bibfield{author}{\bibinfo{person}{F~Maxwell Harper} {and}
  \bibinfo{person}{Joseph~A Konstan}.} \bibinfo{year}{2015}\natexlab{}.
\newblock \showarticletitle{The {MovieLens} datasets: History and context}.
\newblock \bibinfo{journal}{\emph{ACM transactions on interactive intelligent
  systems (TIIS)}} \bibinfo{volume}{5}, \bibinfo{number}{4}
  (\bibinfo{year}{2015}), \bibinfo{pages}{1--19}.
\newblock


\bibitem[Hastie et~al\mbox{.}(2009)]%
        {hastie2009elements}
\bibfield{author}{\bibinfo{person}{Trevor Hastie}, \bibinfo{person}{Robert
  Tibshirani}, \bibinfo{person}{Jerome~H Friedman}, {and}
  \bibinfo{person}{Jerome~H Friedman}.} \bibinfo{year}{2009}\natexlab{}.
\newblock \bibinfo{booktitle}{\emph{The elements of statistical learning: data
  mining, inference, and prediction}}. Vol.~\bibinfo{volume}{2}.
\newblock \bibinfo{publisher}{Springer}.
\newblock


\bibitem[Hu et~al\mbox{.}(2008)]%
        {hu2008collaborative}
\bibfield{author}{\bibinfo{person}{Yifan Hu}, \bibinfo{person}{Yehuda Koren},
  {and} \bibinfo{person}{Chris Volinsky}.} \bibinfo{year}{2008}\natexlab{}.
\newblock \showarticletitle{Collaborative filtering for implicit feedback
  datasets}. In \bibinfo{booktitle}{\emph{2008 Eighth IEEE international
  conference on data mining}}. Ieee, \bibinfo{pages}{263--272}.
\newblock


\bibitem[Ihemelandu and Ekstrand(2021)]%
        {ihemelandu2021statistical}
\bibfield{author}{\bibinfo{person}{Ngozi Ihemelandu} {and}
  \bibinfo{person}{Michael~D. Ekstrand}.} \bibinfo{year}{2021}\natexlab{}.
\newblock \bibinfo{title}{Statistical Inference: The Missing Piece of RecSys
  Experiment Reliability Discourse}.
\newblock
\newblock
\showeprint[arxiv]{2109.06424}~[cs.IR]


\bibitem[Jin et~al\mbox{.}(2021)]%
        {jin2021towards}
\bibfield{author}{\bibinfo{person}{Ruoming Jin}, \bibinfo{person}{Dong Li},
  \bibinfo{person}{Jing Gao}, \bibinfo{person}{Zhi Liu}, \bibinfo{person}{Li
  Chen}, {and} \bibinfo{person}{Yang Zhou}.} \bibinfo{year}{2021}\natexlab{}.
\newblock \showarticletitle{Towards a better understanding of linear models for
  recommendation}. In \bibinfo{booktitle}{\emph{Proceedings of the 27th ACM
  SIGKDD Conference on Knowledge Discovery \& Data Mining}}.
  \bibinfo{pages}{776--785}.
\newblock


\bibitem[Konstan and Adomavicius(2013)]%
        {konstan2013toward}
\bibfield{author}{\bibinfo{person}{Joseph~A Konstan} {and}
  \bibinfo{person}{Gediminas Adomavicius}.} \bibinfo{year}{2013}\natexlab{}.
\newblock \showarticletitle{Toward identification and adoption of best
  practices in algorithmic recommender systems research}. In
  \bibinfo{booktitle}{\emph{Proceedings of the international workshop on
  Reproducibility and replication in recommender systems evaluation}}.
  \bibinfo{pages}{23--28}.
\newblock


\bibitem[Krichene and Rendle(2020)]%
        {krichene2020sampled}
\bibfield{author}{\bibinfo{person}{Walid Krichene} {and}
  \bibinfo{person}{Steffen Rendle}.} \bibinfo{year}{2020}\natexlab{}.
\newblock \showarticletitle{On sampled metrics for item recommendation}. In
  \bibinfo{booktitle}{\emph{Proceedings of the 26th ACM SIGKDD international
  conference on knowledge discovery \& data mining}}.
  \bibinfo{pages}{1748--1757}.
\newblock


\bibitem[Li et~al\mbox{.}(2020)]%
        {li2020sampling}
\bibfield{author}{\bibinfo{person}{Dong Li}, \bibinfo{person}{Ruoming Jin},
  \bibinfo{person}{Jing Gao}, {and} \bibinfo{person}{Zhi Liu}.}
  \bibinfo{year}{2020}\natexlab{}.
\newblock \showarticletitle{On sampling top-k recommendation evaluation}. In
  \bibinfo{booktitle}{\emph{Proceedings of the 26th ACM SIGKDD International
  Conference on Knowledge Discovery \& Data Mining}}.
  \bibinfo{pages}{2114--2124}.
\newblock


\bibitem[Liang et~al\mbox{.}(2018)]%
        {liang2018variational}
\bibfield{author}{\bibinfo{person}{Dawen Liang}, \bibinfo{person}{Rahul~G
  Krishnan}, \bibinfo{person}{Matthew~D Hoffman}, {and} \bibinfo{person}{Tony
  Jebara}.} \bibinfo{year}{2018}\natexlab{}.
\newblock \showarticletitle{Variational autoencoders for collaborative
  filtering}. In \bibinfo{booktitle}{\emph{Proceedings of the 2018 world wide
  web conference}}. \bibinfo{pages}{689--698}.
\newblock


\bibitem[Ludewig et~al\mbox{.}(2019)]%
        {ludewig2019performance}
\bibfield{author}{\bibinfo{person}{Malte Ludewig}, \bibinfo{person}{Noemi
  Mauro}, \bibinfo{person}{Sara Latifi}, {and} \bibinfo{person}{Dietmar
  Jannach}.} \bibinfo{year}{2019}\natexlab{}.
\newblock \showarticletitle{Performance comparison of neural and non-neural
  approaches to session-based recommendation}. In
  \bibinfo{booktitle}{\emph{Proceedings of the 13th ACM conference on
  recommender systems}}. \bibinfo{pages}{462--466}.
\newblock


\bibitem[Marlin(2003)]%
        {marlin2003modeling}
\bibfield{author}{\bibinfo{person}{Benjamin~M Marlin}.}
  \bibinfo{year}{2003}\natexlab{}.
\newblock \showarticletitle{Modeling user rating profiles for collaborative
  filtering}.
\newblock \bibinfo{journal}{\emph{Advances in neural information processing
  systems}}  \bibinfo{volume}{16} (\bibinfo{year}{2003}).
\newblock


\bibitem[Ning and Karypis(2011)]%
        {ning2011slim}
\bibfield{author}{\bibinfo{person}{Xia Ning} {and} \bibinfo{person}{George
  Karypis}.} \bibinfo{year}{2011}\natexlab{}.
\newblock \showarticletitle{Slim: Sparse linear methods for top-n recommender
  systems}. In \bibinfo{booktitle}{\emph{2011 IEEE 11th international
  conference on data mining}}. IEEE, \bibinfo{pages}{497--506}.
\newblock


\bibitem[Paudel et~al\mbox{.}(2016)]%
        {paudel2016updatable}
\bibfield{author}{\bibinfo{person}{Bibek Paudel}, \bibinfo{person}{Fabian
  Christoffel}, \bibinfo{person}{Chris Newell}, {and} \bibinfo{person}{Abraham
  Bernstein}.} \bibinfo{year}{2016}\natexlab{}.
\newblock \showarticletitle{Updatable, accurate, diverse, and scalable
  recommendations for interactive applications}.
\newblock \bibinfo{journal}{\emph{ACM Transactions on Interactive Intelligent
  Systems (TiiS)}} \bibinfo{volume}{7}, \bibinfo{number}{1}
  (\bibinfo{year}{2016}), \bibinfo{pages}{1--34}.
\newblock


\bibitem[Rendle et~al\mbox{.}(2019)]%
        {rendle2019difficulty}
\bibfield{author}{\bibinfo{person}{Steffen Rendle}, \bibinfo{person}{Li Zhang},
  {and} \bibinfo{person}{Yehuda Koren}.} \bibinfo{year}{2019}\natexlab{}.
\newblock \showarticletitle{On the difficulty of evaluating baselines: A study
  on recommender systems}.
\newblock \bibinfo{journal}{\emph{arXiv preprint arXiv:1905.01395}}
  (\bibinfo{year}{2019}).
\newblock


\bibitem[Said and Bellog{\'\i}n(2014)]%
        {said2014comparative}
\bibfield{author}{\bibinfo{person}{Alan Said} {and} \bibinfo{person}{Alejandro
  Bellog{\'\i}n}.} \bibinfo{year}{2014}\natexlab{}.
\newblock \showarticletitle{Comparative recommender system evaluation:
  benchmarking recommendation frameworks}. In
  \bibinfo{booktitle}{\emph{Proceedings of the 8th ACM Conference on
  Recommender systems}}. \bibinfo{pages}{129--136}.
\newblock


\bibitem[Salah et~al\mbox{.}(2020)]%
        {cornac}
\bibfield{author}{\bibinfo{person}{Aghiles Salah}, \bibinfo{person}{Quoc-Tuan
  Truong}, {and} \bibinfo{person}{Hady~W Lauw}.}
  \bibinfo{year}{2020}\natexlab{}.
\newblock \showarticletitle{Cornac: A Comparative Framework for Multimodal
  Recommender Systems}.
\newblock \bibinfo{journal}{\emph{Journal of Machine Learning Research}}
  \bibinfo{volume}{21}, \bibinfo{number}{95} (\bibinfo{year}{2020}),
  \bibinfo{pages}{1--5}.
\newblock


\bibitem[Shani and Gunawardana(2011)]%
        {shani2011evaluating}
\bibfield{author}{\bibinfo{person}{Guy Shani} {and} \bibinfo{person}{Asela
  Gunawardana}.} \bibinfo{year}{2011}\natexlab{}.
\newblock \showarticletitle{Evaluating recommendation systems}.
\newblock In \bibinfo{booktitle}{\emph{Recommender systems handbook}}.
  \bibinfo{publisher}{Springer}, \bibinfo{pages}{257--297}.
\newblock


\bibitem[Steck(2019)]%
        {steck2019embarrassingly}
\bibfield{author}{\bibinfo{person}{Harald Steck}.}
  \bibinfo{year}{2019}\natexlab{}.
\newblock \showarticletitle{Embarrassingly shallow autoencoders for sparse
  data}. In \bibinfo{booktitle}{\emph{The World Wide Web Conference}}.
  \bibinfo{pages}{3251--3257}.
\newblock


\bibitem[Steck et~al\mbox{.}(2020)]%
        {steck2020admm}
\bibfield{author}{\bibinfo{person}{Harald Steck}, \bibinfo{person}{Maria
  Dimakopoulou}, \bibinfo{person}{Nickolai Riabov}, {and} \bibinfo{person}{Tony
  Jebara}.} \bibinfo{year}{2020}\natexlab{}.
\newblock \showarticletitle{{ADMM SLIM}: Sparse recommendations for many
  users}. In \bibinfo{booktitle}{\emph{Proceedings of the 13th International
  Conference on Web Search and Data Mining}}. \bibinfo{pages}{555--563}.
\newblock


\bibitem[Sun et~al\mbox{.}(2020)]%
        {sun2020we}
\bibfield{author}{\bibinfo{person}{Zhu Sun}, \bibinfo{person}{Di Yu},
  \bibinfo{person}{Hui Fang}, \bibinfo{person}{Jie Yang},
  \bibinfo{person}{Xinghua Qu}, \bibinfo{person}{Jie Zhang}, {and}
  \bibinfo{person}{Cong Geng}.} \bibinfo{year}{2020}\natexlab{}.
\newblock \showarticletitle{Are we evaluating rigorously? benchmarking
  recommendation for reproducible evaluation and fair comparison}. In
  \bibinfo{booktitle}{\emph{Fourteenth ACM conference on recommender systems}}.
  \bibinfo{pages}{23--32}.
\newblock


\bibitem[Sutskever et~al\mbox{.}(2013)]%
        {sutskever2013importance}
\bibfield{author}{\bibinfo{person}{Ilya Sutskever}, \bibinfo{person}{James
  Martens}, \bibinfo{person}{George Dahl}, {and} \bibinfo{person}{Geoffrey
  Hinton}.} \bibinfo{year}{2013}\natexlab{}.
\newblock \showarticletitle{On the importance of initialization and momentum in
  deep learning}. In \bibinfo{booktitle}{\emph{International conference on
  machine learning}}. PMLR, \bibinfo{pages}{1139--1147}.
\newblock


\bibitem[Tamm et~al\mbox{.}(2021)]%
        {qualitymetrics}
\bibfield{author}{\bibinfo{person}{Yan-Martin Tamm}, \bibinfo{person}{Rinchin
  Damdinov}, {and} \bibinfo{person}{Alexey Vasilev}.}
  \bibinfo{year}{2021}\natexlab{}.
\newblock \bibinfo{booktitle}{\emph{Quality Metrics in Recommender Systems: Do
  We Calculate Metrics Consistently?}}
\newblock \bibinfo{publisher}{Association for Computing Machinery},
  \bibinfo{address}{New York, NY, USA}, \bibinfo{pages}{708–713}.
\newblock
\showISBNx{9781450384582}
\urldef\tempurl%
\url{https://doi.org/10.1145/3460231.3478848}
\showURL{%
\tempurl}


\bibitem[Vapnik(1999)]%
        {vapnik1999nature}
\bibfield{author}{\bibinfo{person}{Vladimir Vapnik}.}
  \bibinfo{year}{1999}\natexlab{}.
\newblock \bibinfo{booktitle}{\emph{The nature of statistical learning
  theory}}.
\newblock \bibinfo{publisher}{Springer science \& business media}.
\newblock


\bibitem[Zhao et~al\mbox{.}(2020)]%
        {recbole}
\bibfield{author}{\bibinfo{person}{Wayne~Xin Zhao}, \bibinfo{person}{Shanlei
  Mu}, \bibinfo{person}{Yupeng Hou}, \bibinfo{person}{Zihan Lin},
  \bibinfo{person}{Kaiyuan Li}, \bibinfo{person}{Yushuo Chen},
  \bibinfo{person}{Yujie Lu}, \bibinfo{person}{Hui Wang},
  \bibinfo{person}{Changxin Tian}, \bibinfo{person}{Xingyu Pan},
  \bibinfo{person}{Yingqian Min}, \bibinfo{person}{Zhichao Feng},
  \bibinfo{person}{Xinyan Fan}, \bibinfo{person}{Xu Chen},
  \bibinfo{person}{Pengfei Wang}, \bibinfo{person}{Wendi Ji},
  \bibinfo{person}{Yaliang Li}, \bibinfo{person}{Xiaoling Wang}, {and}
  \bibinfo{person}{Ji-Rong Wen}.} \bibinfo{year}{2020}\natexlab{}.
\newblock \showarticletitle{RecBole: Towards a Unified, Comprehensive and
  Efficient Framework for Recommendation Algorithms}.
\newblock \bibinfo{journal}{\emph{arXiv preprint arXiv:2011.01731}}
  (\bibinfo{year}{2020}).
\newblock


\end{thebibliography}

\end{document}


\appendix

\section{Hyperparameter Ranges}
\begin{itemize}
\item ALS
\begin{itemize}
\item \verb+n_factors+: Discrete with range/values: (16, 512)
\item \verb+regularization+: Continuous with range/values: (1e-06, 200)
\end{itemize}
\item Ease
\begin{itemize}
\item \verb+lmbda+: Continuous with range/values: (1, 1000)
\end{itemize}
\item MultiDAE
\begin{itemize}
\item \verb+p_dims+: Categorical with range/values: ([200, 'n\_items'], [200, 600, 'n\_items'], [200, 600, 600, 'n\_items'])
\item \verb+lr+: Categorical with range/values: (0.0001, 0.001, 0.01)
\item \verb+lam+: LogContinuous with range/values: (1e-06, 10.0)
\end{itemize}
\item MultiVAE
\begin{itemize}
\item \verb+p_dims+: Categorical with range/values+: ([200, 'n\_items'], [200, 600, 'n\_items'], [200, 600, 600, 'n\_items'])
\item \verb+lr+: Categorical with range/values+: (0.0001, 0.001, 0.01)
\item \verb+anneal_cap+: Continuous with range/values+: (0.01, 1.0)
\end{itemize}
\item P3alpha
\begin{itemize}
\item \verb+normalize_similarity+ Categorical with range/values+: [True, False]
\item \verb+alpha+ Continuous with range/values+: (0.0, 2.0)
\item \verb+k+: Discrete with range/values: (5, 1000)
\end{itemize}
\item \verb+Popularity+: n/a
\item PureSVD
\begin{itemize}
\item \verb+n_factors+: Discrete with range/values: (16, 512)
\end{itemize}
\item RP3beta
\begin{itemize}
\item \verb+normalize_similarity+: Categorical with range/values: [True, False]
\item \verb+alpha+: Continuous with range/values: (0.0, 2.0)
\item \verb+beta+: Continuous with range/values: (0.0, 2.0)
\item \verb+k+: Discrete with range/values: (5, 1000)
\end{itemize}
\item \verb+Random+: n/a
\item SLIM
\begin{itemize}
\item \verb+l1+: LogContinuous with range/values: (0.1, 10000)
\item \verb+l2+: LogContinuous with range/values: (0.1, 10000)
\end{itemize}
\item UserKNN
\begin{itemize}
\item \verb+weighting+: Categorical with range/values: ('uniform', 'similarity')
\item \verb+k+: Discrete with range/values: (1, 500)
\item \verb+lambda+: LogContinuous with range/values: (0.1, 1000.0)
\item \verb+alpha+: Continuous with range/values: (0.1, 0.9)
\end{itemize}
\item ItemKNN
\begin{itemize}
\item \verb+weighting+: Categorical with range/values: ('uniform', 'similarity')
\item \verb+k+: Discrete with range/values: (1, 500)
\item \verb+lambda+: LogContinuous with range/values: (0.1, 1000.0)
\item \verb+alpha+: Continuous with range/values: (0.1, 0.9)
\end{itemize}
\end{itemize}